\documentclass{article}
\usepackage{graphicx}  
\usepackage{amsmath}   
\usepackage[compress]{cite}
\usepackage{amssymb}   
\usepackage{bm} 
\usepackage{dcolumn}
\usepackage{color}
\usepackage{mathrsfs}
\usepackage{amsfonts}
\usepackage{varioref}
\usepackage[active]{srcltx}
\usepackage{tikz}
\usepackage{lmodern}
\usetikzlibrary{decorations.pathmorphing}
\tikzset{snake it/.style={decorate, decoration=snake}}
\usepackage{fancybox}
\RequirePackage[colorlinks,citecolor=blue,urlcolor=magenta,linkcolor=blue]{hyperref}
\def\EH{Einstein-Hilbert }

\def\gr{general relativity}
\def\RN{Reissner-Nordstr\"{o}m }

\def\KR{Kalb-Ramond }
\labelformat{section}{Section #1} 
\labelformat{subsection}{Section #1} 
\labelformat{subsubsection}{Section #1}
\labelformat{subsubsubsection}{Section #1}
\labelformat{equation}{Eq.~(#1)} 
\labelformat{figure}{Fig.~#1} 
\labelformat{subfigure}{Fig.~\thefigure#1} 
\labelformat{table}{Table~#1} 
\labelformat{appendix}{Appendix #1}

\title{Strong gravitational lensing --- A probe for extra dimensions and Kalb-Ramond field}

\author{Sumanta Chakraborty \footnote{sumantac.physics@gmail.com} 
and
Soumitra SenGupta \footnote{tpssg@iacs.res.in}\\
{\small{Department of Theoretical Physics, Indian Association for the Cultivation of Science, Kolkata-700032, India}}}

\begin{document}
  
\maketitle
\begin{abstract}
Strong field gravitational lensing in the context of both higher spacetime dimensions and in presence of Kalb-Ramond field have been studied. After developing proper analytical tools to analyze the problem we consider gravitational lensing in three distinct black hole spacetimes --- (a) four dimensional black hole in presence of Kalb-Ramond field, (b) brane world black holes with Kalb-Ramond field and finally (c) black hole solution in $f(T)$ gravity. In all the three situations we have depicted the behavior of three observables: the asymptotic position approached by the relativistic images, the angular separation and magnitude difference between the outermost images with others packed inner ones, both numerically and analytically. Difference between these scenarios have also been discussed along with possible observational signatures.
\end{abstract}
\section{Introduction}

Gravitational lensing is one of the most routinely used astrophysical observations to infer gravitational mass of distant galaxies or galaxy clusters. In a broad sense, gravitational lensing corresponds to deflection of electromagnetic waves from its original trajectory \'{a} la spacetime curvature. Since spacetime curvature depends on the gravitational mass of the body producing it, the measurement of the deflection angle of the electromagnetic wave will be related to the mass of the gravitating object. Actually, gravitational lensing or more commonly known as the bending of light, in the context of solar system was used as the first observational verification of Einstein's general relativity. However all the situations depicted above correspond to weak field limit, where the spacetime curvature experienced by the electromagnetic wave is small such that the deflection angle is much small compared to $2\pi$ (see e.g., \cite{Schneider1992,Liebes:1964zz,Refsdal:1993kf} and the references therein). 

On the other hand, there can be situations where the deflection angle can become pretty large, e.g., when an electromagnetic wave passes close to a black hole, it can actually hover around the hole in closed loops before actually escaping it. In particular, there exist a sphere across the black hole, known as the photon sphere, where the deflection angle will even diverge. Further, the strong field lensing develops an infinite set of discrete images near the central massive object, known as relativistic images, having no analogue in the context of weak gravitational lensing. The idea that black holes can act as gravitational lens was first put forward by Darwin in \cite{Darwin:1959}, which subsequently was elaborated in \cite{Ohanian:1987,Luminet:1979}. However the current theoretical formulation of strong gravitational lensing from a black hole can be attributed to the seminal work of Virbhadra and Ellis, where existence of such relativistic images were predicted in the context of 
Schwarzschild 
spacetime \cite{Virbhadra:1999nm}. Immediately afterwards the concepts laid down in \cite{Virbhadra:1999nm} were formulated in a more rigorous way by properly defining the strong field limit and providing better analytical 
understanding of the problem \cite{Frittelli:1999yf,Bozza:2001xd}. Subsequently it has been further generalized by providing the analytical framework in a general spherically symmetric spacetime \cite{Bozza:2002zj,Bozza:2007gt} and has been applied to various black hole spacetimes which includes --- Reissner-Nordstr\"{o}m \cite{Eiroa:2002mk}, Kerr \cite{Bozza:2006nm,Kraniotis:2014paa}, Einstein-Born-Infeld \cite{Eiroa:2005ag}, Horava-Lifshitz \cite{Chen:2009eu}, Brane world scenarios \cite{Whisker:2004gq}, dilaton \cite{Bhadra:2003zs,Ghosh:2010uw,Mukherjee:2006ru}, phantom \cite{Eiroa:2013nra,Gyulchev:2012ty} and Galileon \cite{Zhao:2016kft}. It has also been applied to wormholes \cite{Nandi:2006ds}, regular black holes \cite{Wei:2015qca}, self-dual black holes in loop quantum gravity \cite{Sahu:2015dea} and naked singular spacetimes \cite{Virbhadra:2002ju,Sahu:2012er}. Also see \cite{Amore:2006mc,Bozza:2009yw,Raffaelli:2014ola} for recent reviews. 

The growing interest in the strong gravitational lensing is mainly due to the reason that one can use the lensing phenomenon to infer properties about the regions in the vicinity of the black hole horizon. In particular through gravitational lensing one can infer the region of black hole spacetime, known as black hole shadows. Current observations cannot access these shadows mainly due to low resolution, but in upcoming years with event horizon telescope, very long baseline interferometry, square kilometer array one would be able to resolve regions very near the black hole where the strong field gravitational lensing will become very important \cite{Loeb:2013lfa,Bozza:2004kq,Abdujabbarov:2015xqa,Falcke:2013ola,Bozza:2012by,BinNun:2010ty,Bozza:2015wbw,Bozza:2007gm,
Bozza:2014ywa,Bozza:2015haa,Aldi:2016ntn}. 

Given the current theoretical importance and future observational backdrops it is important to study some fundamental physics using strong gravitational lensing. Two such important aspects are presence of extra spacetime dimensions and \KR field or spacetime torsion. In the higher dimensional physics the standard scenario corresponds to matter fields confined to the brane, while gravity can exist in the bulk. In string theoretic description there exists a closed string excitation leading to a second rank antisymmetric tensor field, known as the \KR field. Thus if the \KR field exists then it must invade the extra dimensions \cite{Green:1987mn}, suggesting that one has to employ the Gauss-Codazzi formalism to obtain the effective theory on the four dimensional hypersurface \cite{Chakraborty:2014xla,Chakraborty:2015bja,Chakraborty:2015taq,Chakraborty:2016ydo,gravitation}. This antisymmetric tensor plays an important role in many areas of physics, which includes, e.g., theories of gravity formulated using 
twistors 
requires the \KR field \cite{Howe:1996kj}, spacetime with \KR field can become optically active \cite{Kar:2001eb}, \KR field can give rise to topological defects, which might lead to intrinsic angular momentum to the structures in our galaxies \cite{Letelier:1995ze}, \KR field also plays an important role in understanding leptogenesis \cite{Ellis:2013gca}, \KR field also affects the anisotropy in cosmic microwave background \cite{Maity:2004he} and so on. In particular given the second rank antisymmetric \KR field one can construct a third rank antisymmetric tensor, which appears naturally in the gravity theory and can act as the missing part of \gr, the spacetime torsion \cite{Majumdar:1999jd}. However in this work we will \emph{not} identify spacetime torsion as originating from the \KR field, but shall treat them on distinct footing. With \KR field we will discuss two situations --- (i) \KR field in four spacetime dimensions and (ii) \KR field exists in higher dimensions (referred to as the bulk) and its 
effect on a lower dimensional hypersurface (known as the brane). Then we separately discuss the scenario with spacetime torsion in the context of $f(T)$ gravity, which has interesting cosmological implications \cite{Myrzakulov:2010vz,Chen:2010va,Dent:2011zz,Bamba:2010wb,Zhang:2011qp,Chakraborty:2012kj,Yang:2010ji}. In this work we derive the gravitational lens equation and the deflection suffered by the light ray. Finally corresponding observables have also been introduced in the context of \KR field in four and in higher spacetime dimensions as well as in the presence of spacetime torsion in the form of $f(T)$ gravity. 

The paper is organized as follows: In \ref{Sec_Analytical} we present necessary analytical results in a compact manner for the most general static, spherically symmetric solution. The same is subsequently applied in \ref{Lensing_Calc} to three different situations --- (a) \KR field in four spacetime dimensions (discussed in \ref{KRfour}), (b) \KR field in higher dimensions (illustrated in \ref{KRfive}) and finally (c) spacetime torsion in the context of $f(T)$ gravity (depicted in \ref{fTgrav}). We end with a discussion on the results obtained. Detailed calculations are presented in \ref{App_01} and \ref{App_02} respectively. 
\section{Strong gravitational lensing: Analytical results}\label{Sec_Analytical}

Static and spherically symmetric spacetimes form an integral part of black hole physics where all the basic features of black holes can be demonstrated and can further be used to understand and hence generalize them as fit for more general scenarios. In this work as well, we will content ourselves with static and spherically symmetric spacetimes but in the presence of Kalb-Ramond field and extra dimensions. We will first summarize the key concepts involved and the observables related to strong gravitational lensing for a general static, spherically symmetric spacetime, following the seminal work by Bozza \cite{Bozza:2002zj} before considering the spacetimes of our interest. 

\begin{figure}
        \begin{tikzpicture}[scale=0.9]   
            \draw[cyan!80,dotted,line width=0.5mm] (-8,0) -- (8,0);
            \draw[blue!90,line width=0.5mm] (0,0) circle (1cm);
            \fill[blue!100] (0,0) circle (0.4 cm);
            \draw[blue!90,line width=0.5mm] (-8,0) .. controls (-1,1.2) and (1,1.2) .. (8,0.8);
            \draw[magenta!80,dashdotted,line width=0.5mm] (-8,0) -- (8,0.8);
            \draw[magenta!80,dashdotted,line width=0.5mm] (-8,0) -- (8,3); 
            \draw[cyan!80,dotted,line width=0.5mm] (8,0) -- (8,3);   
            \draw[black!80,dotted,line width=0.5mm] (0,0) -- (0,1.5);  
            \draw[black!80,dotted,line width=0.5mm] (0,1.5) -- (8,0.8);
            \draw[->,black!80,dotted,line width=0.5mm] (0,0) -- (-0.4,1.5);
            \fill[red!100] (8,0.8) circle (0.1 cm);
            \fill[red!100] (8,3) circle (0.1 cm);
            \fill[red!100] (-8,0) circle (0.1 cm);
            \node[label=right:\textrm{Source}] at (8.1,0.8) {};
            \node[label=left:\textrm{Lens}] at (0.6,-0.6) {};
            \node[label=right:\textrm{Image}] at (8.1,3) {};
            \node[label=right:\textrm{Observer}] at (-8.3,-0.3) {};
            \node[label=below:$\theta$] at (-6,0.55) {};
            \node[label=below:$\beta$] at (3,0.7) {};
	    \node[label=below:$\Theta$] at (3.0,2.0) {};
	    \node[label=left:$u$] at (-0.25,1.15) {};
	    \node[label=above:$D_{\rm OL}$] at (-4,-2) {};
	    \node[label=above:$D_{\rm LS}$] at (4,-2) {};
	    \node[label=above:$D_{\rm OS}$] at (0,-3) {};
            \draw[<->] (-8,-2) to (0,-2);
            \draw[<->] (8,-2) to (0,-2);
            \draw[<->] (-8,-3) to (8,-3);
            \draw[->,bend right=35] (2.5,1.3) to (2.5,2.0); 
            \draw[->,bend right=35] (2.5,0) to (2.5,0.5); 
            \draw[->,bend right=35] (-6.5,0) to (-6.6,0.3);        
        \end{tikzpicture}
    \caption{Schematic diagram of strong gravitational lensing. The observer, the source, the lens and the image are depicted. The light emitted by the source (thick blue line) gets deflected by the lens and reaches the observer. From the viewpoint of the observer it will appear that the light ray is originating from the image. Three different angular separations are also shown, $\beta$ being the angular separation between the source and the lens, $\theta$ stands for the angular difference between the image and the lens while $\Theta$ representing the deflection. Three distances $D_{\rm OL}$, $D_{\rm LS}$ and $D_{OS}$ representing the observer-lens, lens-source and observer-source have also been shown.}
    \label{Lensing_Schematic_01}
\end{figure}
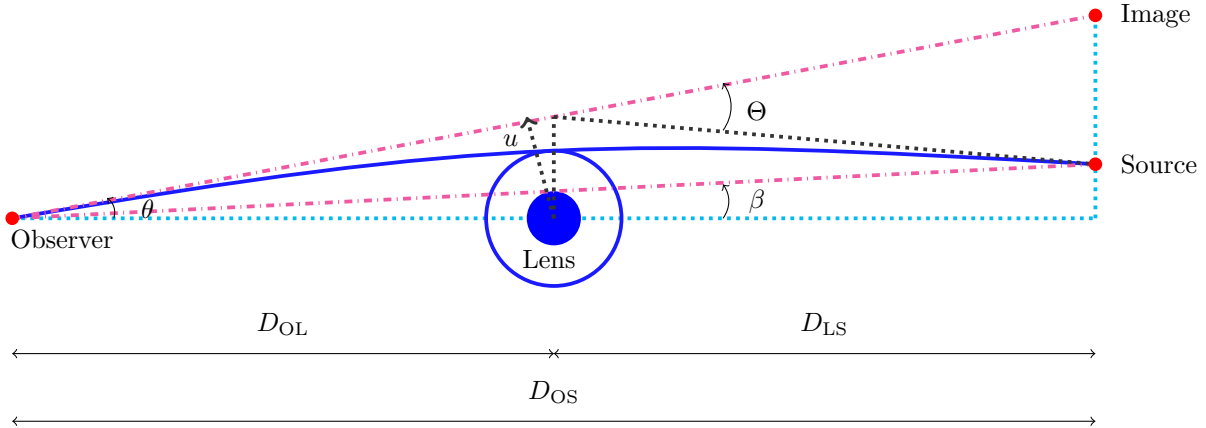

For this purpose, let us start with a static, spherically symmetric spacetime described by the following line element,
\begin{align}\label{Eq_Lens_01}
ds^{2}=-f(r)dt^{2}+\frac{dr^{2}}{g(r)}+r^{2}d\Omega _{2}^{2}
\end{align}
where, $f(r)$ and $g(r)$ are arbitrary functions of the radial coordinate $r$ with $d\Omega _{2}^{2}$ being the line element on  two sphere. (Some interesting features regarding this line element have been explored in \cite{Bhattacharya:2016naa,Chakraborty:2011uj,Chakraborty:2014eha,Chakraborty:2015vla}) We further assume that the spacetime is asymptotically flat, i.e., $f(r\rightarrow \infty)=1=g(r\rightarrow \infty)$. We will be exclusively considering the trajectories of photons in this work and the single most important radius associated with those trajectories correspond to the radius of photon circular orbit $r_{\rm ph}$. This can be obtained by solving the following algebraic equation,
\begin{align}\label{Eq_Lens_02}
rf'(r)-2f(r)=0
\end{align}
where $f'(r)$ stands for derivative of the metric function $f(r)$ with respect to radial coordinate $r$. Note that the above equation is independent of the structure of $g(r)$, i.e., the $g_{tt}$ component is solely responsible for the structure of the photon sphere. The trajectory of a photon is specified by two constants of motion, the energy $E$ and the angular momentum $L$. In terms of these two constants of motion the trajectory of the photon can be represented by the following differential equation,
\begin{align}\label{Eq_Lens_03}
\frac{d\phi}{dr}=\frac{L}{r^{2}}\sqrt{\frac{f}{g}}\left[E^{2}-f(r)\frac{L^{2}}{r^{2}}\right]^{-1/2}
\end{align}
The minimum radius $r_{0}$ that a photon can travel with energy $E$ and angular momentum $L$ corresponds to the one when value of the term inside square root vanishes. Thus the impact parameter $b=L/E$ can be written in terms of $r_{0}$ as, $(1/b^{2})=(f(r_{0})/r_{0}^{2})$ (see \ref{Lensing_Schematic_01} as well as \cite{Bozza:2002zj} for details). 
Hence the orbit equation gets modified to,
\begin{align}\label{Eq_Lens_04}
\frac{d\phi}{dr}=\frac{1}{r^{2}}\sqrt{\frac{f}{g}}\left[\frac{f(r_{0})}{r_{0}^{2}}-\frac{f(r)}{r^{2}}\right]^{-1/2}
\end{align}
Integrating this orbit equation one obtains the deflection angle suffered by the photon as it travels from asymptotic infinity to the impact distance $r_{0}$ as,
\begin{align}\label{Eq_Lens_05}
\Theta &=-\pi +\int _{x_{0}}^{x_{\infty}}dx~R(x,r_{0})F(x,r_{0})
\nonumber
\\
R(x,r_{0})&=2\frac{dr}{dx}\sqrt{\frac{f}{g}}\frac{r_{0}}{r^{2}};\qquad F(x,r_{0})=\left[f(r_{0})-\frac{r_{0}^{2}}{r^{2}}f(r)\right]^{-1/2}
\end{align}
In the above integral we have introduced a new variable $x=x(r)$, yet unspecified, such that $x(r_{0})=x_{0}$ and $x(\infty)=x_{\infty}$. A standard choice is to assume $x=(f(r)-f(r_{0}))/(1-f(r_{0}))$, such that as $r\rightarrow \infty$ one obtains $x_{\infty}=1$, while for $r\rightarrow r_{0}$ one arrives at $x_{0}=0$. However we will not assume any such particular expression for $x$ and will keep it general for the moment, only with $x_{0}=0$. The reason for separating out the integral in the deflection angle into two parts is immediate, while $R(x,r_{0})$ is finite for all values of $r$, the other term $F(r_{0},x)$ diverges as the photon approaches $r=r_{0}$. To deal with the divergence, one can Taylor expand the function $f(r_{0})-(r_{0}^{2}f(r)/r^{2})$ inside the inverse square root around $r=r_{0}$. This essentially implies expansion around $x=0$. Performing the expansion, one can separate out the integral into a regular part and a divergent part (for more detailed discussion see \cite{Bozza:2002zj}). 
The divergent part can further be integrated analytically and expanded around $r=r_{\rm ph}$ (see \ref{App_01} for detailed derivation). 

For convenience, one can introduce another variable $u=r/\sqrt{f(r)}$ as a substitute for the radial coordinate $r$, such that $u_{0}$ leads to the impact parameter of the photon. Then rewriting the divergent integral in terms of the new variable $u$, we finally obtain the expression for deflection angle to read \cite{Bozza:2002zj},
\begin{align}\label{Lens_Main}
\Theta (\theta)=-\bar{a}\ln \left(\frac{\theta D_{\rm OL}}{u_{\rm ph}}-1\right)+\bar{b} 
\end{align}
where, $D_{\rm OL}$ stands for the distance between the observer and the lens, $\theta$ being the angle between the image and the lens and $u_{\rm ph}=r_{\rm ph}/\sqrt{f(r_{\rm ph})}$ (see \ref{Lensing_Schematic_01}). The other objects appearing in the above expression has the following structure in terms of the functions $f(r)$, $g(r)$ and their derivatives,
\begin{align}
\bar{a}&=\sqrt{\frac{2f(r_{\rm ph})}{g(r_{\rm ph})\left[-r_{\rm ph}^{2}f''(r_{\rm ph})+2f(r_{\rm ph})\right]}}
\label{Eq_Lens_06a}
\\
\bar{b}&=-\pi +b_{\rm R}+\bar{a}\ln \left(\frac{\left[-r_{\rm ph}^{2}f''(r_{\rm ph})+2f(r_{\rm ph})\right]\left[1-f(r_{\rm ph})\right]^{2}}{r_{\rm ph}^{2}f(r_{\rm ph})\left[f'(r_{\rm ph})\right]^{2}}\right)
\label{Eq_Lens_06b}
\\
b_{\rm R}&=\int _{0}^{1}dx \Bigg\lbrace \frac{2r_{\rm ph}}{r(x)^{2}}\sqrt{\frac{f(x)}{g(x)}}\left( \frac{1-f(x)}{f'(x)} \right)\left[f(r_{\rm ph})-\frac{r_{\rm ph}^{2}}{r(x)^{2}}f(x)\right]^{-1/2}-\frac{2\bar{a}}{x}\Bigg\rbrace
\label{Eq_Lens_06c}
\end{align}
where, we have used the standard expression connecting $x$ and $r$ to be, $x=(f(r)-f(r_{\rm ph}))(1-f(r_{\rm ph}))$. With all these identifications \ref{Lens_Main} yields the general deflection angle for gravitational lensing in the strong field regime (see also \cite{Bozza:2002zj}). The corresponding observables are --- (i) the deflection angle $\theta _{\infty}$ corresponding to the innermost image, (ii) the angular separation $s$ between the outermost image and the inner ones and finally (iii) the astronomical magnitude difference $r$ between the outermost image and the inner images. These three observables can be expressed in terms of the variables introduced above through the use of the lens equation. We will adopt the lens equation, which was first used and elaborated in \cite{Bozza:2001xd} before relating the above observables to corresponding theoretical expressions. 

One assumes quiet naturally, that the source is places behind the lens. Then assuming that both the source and the observer are asymptotically far away, one can invoke first order approximations to obtain the lens equation to read \cite{Bozza:2001xd} (see also \ref{Lensing_Schematic_01}),
\begin{align}\label{Lens_Eq}
\beta =\theta -\frac{D_{LS}}{D_{OS}}\Delta \Theta _{n}
\end{align}
Here, $\beta$ is the angular separation between the source and the lens, $\theta$ being the angular separation between the image and the lens (see \ref{Lensing_Schematic_01}), while $\Delta \Theta _{n}\equiv \Theta (\theta)-2n\pi$ represents the deflection angle as a photon loops around the hole $n$ times before escaping. The other two parameters appearing in \ref{Lens_Eq} are simply the distance between the source to the lens and the distance between the observer and the source respectively, along the optical axis (see also \cite{Bozza:2008ev} for an improved version of the above lens equation).

Among others one of the most important aspect of the lens equation (see \ref{Lens_Eq}) stems from the fact that it connects the true position of the source (represented by $\beta$) with the apparent position of the image $\theta$. The other parameter, namely $\Theta(\theta)$ can be computed as depicted in \ref{App_01} and can be connected to the background spacetime \cite{Bozza:2002zj}. Thus one can use the theoretical deflection angle expressed as \ref{Lens_Main} in the lens equation to obtain various lensing observables in terms of the corresponding theoretical counterparts. 

For completeness, we will briefly recall how the same can be performed, however for a detailed discussion we refer our reader to \cite{Bozza:2002zj}. As a first step one tries to derive the angular separation $\theta _{n}$, which corresponds to a deflection angle of $2n\pi$ and thus is a solution of the equation $\Theta (\theta _{n})=2n\pi$ \cite{Bozza:2002zj}:
\begin{align}\label{Eq_Rev_01}
\theta _{n}=\frac{u_{\rm ph}}{D_{\rm OL}}\left[1+\exp\left(\frac{\bar{b}-2n\pi}{\bar{a}}\right)\right]
\end{align}
Then one can Taylor expand $\Theta(\theta)$ around $\theta _{n}$ and hence compute $\Theta (\theta)-2n\pi$, which will be proportional to $\theta -\theta _{n}$. The expression for $\Theta (\theta)-2n\pi$ so obtained can subsequently be substituted in \ref{Lens_Eq} leading to \cite{Bozza:2002zj},
\begin{align}
\beta =\theta _{n}+\left(\theta -\theta _{n}\right)+\left(\frac{\bar{a}D_{\rm OL}D_{\rm LS}}{u_{\rm ph}D_{\rm OS}}\right)\exp\left(-\frac{\bar{b}-2n\pi}{\bar{a}}\right)\left(\theta -\theta _{n}\right)
\end{align}
One can invert this equation and hence obtain an expression for $\theta-\theta _{n}$, which provides the angular separation between various relativistic images formed due to gravitational lensing and yields important information about the lensing object, which could be a black hole. Even though in principle the angular separation between any two lensed images is detectable, but for practical purposes one concentrates on the angular separation between the outermost image $\theta _{1}$ and the innermost one $\theta _{\infty}$ (obtained by taking the limit $n\rightarrow \infty$ in \ref{Eq_Rev_01}). Thus an observable associated with gravitational lensing corresponds to, the angular separation $s=\theta _{1}-\theta _{\infty}$ \cite{Bozza:2002zj}. In order to define the other observable it is instructive to introduce the magnification of lensed images. For the n-th image the magnification reads \cite{Bozza:2002zj},
\begin{align}
\mu _{n}=\frac{1}{\left(\beta/\theta\right)\partial \beta/\partial \theta}\vert_{\theta _{n}}
=\frac{u_{\rm ph}^{2}D_{\rm OS}}{\bar{a}\beta D_{\rm OL}^{2}D_{\rm LS}}\exp\left(\frac{\bar{b}-2n\pi}{\bar{a}}\right)\left[1+\exp\left(\frac{\bar{b}-2n\pi}{\bar{a}}\right)\right]
\end{align}
One can immediately understand the importance of this relation, which relates an observable (in this case the magnification) to various theoretical parameters using the strong field lensing. However, since all the lensed images except the $n=1$ are of lesser prominence, one often considers the bolometric magnitude $r$, obtained through the ratio of $\mu _{1}$ and the rest of the lensed images. To simplify the algebra one often notes that since both $\bar{a}$ and $\bar{b}$ are of order unity, it immediately follows that, $\exp(2\pi/\bar{a})\gg1$, as well as $\exp(\bar{b}/\bar{a})\sim 1$. Use of these relations finally leads us the following expression for the lensing observables \cite{Bozza:2002zj}
\begin{align}
\theta _{\infty}&=\frac{u_{\rm ph}}{D_{\rm OL}},
\label{Eq_Lens_07a}
\\
s&=\theta _{1}-\theta _{\infty}=\theta _{\infty}\exp \left(\frac{\bar{b}}{\bar{a}}-\frac{2\pi}{\bar{a}} \right)
\label{Eq_Lens_07b}
\\
r&=2.5 \textrm{log}_{10}\left(\frac{\mu_{1}}{\sum _{n=2}^{\infty}\mu _{n}}\right)=\frac{5\pi}{\bar{a}\log 10}
\label{Eq_Lens_07c}
\end{align}
Thus there is a one to one mapping between the observables $(\theta _{\infty},s,r)$ and the parameters of the model, namely, $(u_{\rm ph},\bar{a},\bar{b})$. Hence given the metric functions $f(r)$ and $g(r)$ one can compute the photon radius and hence each terms in the deflection angle as well as in the lens equation. Then one can numerically compute the above observables and study their difference with the corresponding Schwarzschild one. This will hint towards possible modifications in the behavior of gravitational interactions in the strong field regime and hence have the potential to open up new avenues of exploration. In the next sections we will study the strong field lensing for three black hole spacetimes in presence of Kalb-Ramond field and shall discuss possible departure from the Schwarzschild scenario which will provide a characteristic signature of these spacetimes. For our later convenience we will choose $GM=1=c$ and we shall write the later expressions accordingly.
\section{Lensing in presence of Kalb-Ramond field and extra dimensions}\label{Lensing_Calc}

In this section, we will discuss three situations involving Kalb-Ramond field as well as extra spatial dimensions. In the first example, we mainly concentrate on the effect of Kalb-Ramond field in four dimensional static spherically symmetric spacetimes and the resulting modifications of the strong gravitational lensing observables. Then we introduce additional spatial dimensions and their possible signatures in the effective four-dimensional spacetime. This, in addition to Kalb-Ramond field the effects originating from the bulk spacetime will also be addressed. Finally, we concentrate on a spherically symmetric and static solution in $f(T)$ theories of gravity. We will now elaborate on these different scenario.
\subsection{Kalb-Ramond field in four spacetime dimensions and gravitational lensing}\label{KRfour}

Let us consider a situation when the \KR field is present in four spacetime dimensions. The corresponding gravitational action would involve the \EH term, i.e., the Ricci scalar and the term $\sqrt{-g}H_{abc}H^{abc}$, where $H_{abc}$ is the Kalb-Ramond field. Our interest lies in the static, spherically symmetric solution with this particular modifications of the \EH action. It will turn out that with these enhanced symmetries the \KR field has a single non-trivial spherically symmetric component, which to leading order behaves as, $(2/r^{2})\sqrt{(b/24\pi G_{N})}$ \cite{Kar:2002xa,SenGupta:2001cs}. Thus the parameter $b$ captures the existence of \KR field, since for $b=0$, the \KR field identically vanishes. Solving the gravitational field equations, one obtains the following behaviour of the metric functions \cite{Kar:2002xa},
\begin{align}
f(r)&=1-\frac{2}{r}-\frac{b}{3r^{3}}\left\lbrace 1+\frac{2}{r}+\frac{18}{5r^{2}}+\mathcal{O}\left(\frac{1}{r^{3}}\right)\right\rbrace
+\mathcal{O}\left(\frac{b^{2}}{r^{5}}\right)
\\
g(r)&=1-\frac{2}{r}-\frac{b}{r^{2}}\left\lbrace 1+\frac{1}{r}+\frac{4}{3r^{2}}+\mathcal{O}\left(\frac{1}{r^{3}}\right)\right\rbrace
+\mathcal{O}\left(\frac{b^{2}}{r^{5}}\right)
\end{align}
where we have set $GM=1$ as stated earlier. Note that if the \KR parameter $b$ vanishes, the above metric elements reduce to that of Schwarzschild metric. Given the above metric elements the location of photon circular orbit can be obtained by solving the following algebraic equation,
\begin{align}\label{Eq_Lens_Rev_01}
2=\frac{6}{r}+\frac{b}{r^{3}}\left\lbrace \frac{5}{3}+\frac{4}{r}+\frac{42}{5r^{2}}+\mathcal{O}\left(\frac{1}{r^{3}} \right)\right\rbrace +\mathcal{O}\left(\frac{b^{2}}{r^{5}}\right)
\end{align}
Note that when $b=0$, the above equation leads to the solution $r=3$, representing photon circular orbit in Schwarzschild spacetime with $GM=1$ units. For $b\neq 0$, the leading order contribution comes from the term $(5b/3r^{3})$. If one keeps higher powers of $(1/r)$ as well as that of $(b/r)$, the above equation will lead to very complicated analytical expressions for the photon circular orbit. This stems from the fact that a fully consistent strong field treatment of the above metric will involve an expansion in the powers of $(2/r)$, while we are evaluating the same on the photon circular orbit located near $r=3$. It is obvious that in order to get an accurate result, it is necessary to keep very high order terms in the expansion. Thus in order to provide analytical expressions to the lensing observables in terms of the \KR parameter $b$ we will work exclusively to the leading order term in $(b/r^{3})$. This will capture all the effects of the \KR field, while keeping the mathematics 
simple. However it is necessary to demonstrate the effect of higher order terms in the metric elements on the lensing observables as well. For that purpose we will perform a numerical analysis of the complete situation with metric elements depending on higher order terms.
\begin{table}
\caption{The observables associated with strong gravitational lensing, namely $(\theta _{\infty},s,r)$ and the theoretical parameters of the model, i.e., $(u_{\rm ph},\bar{a},\bar{b})$ have been presented. Numerical estimates of these six quantities for the Schwarzschild black hole as well as for the \RN black hole and the \KR black hole for three possible values of Kalb-Ramond parameter and electric charge have been illustrated. In all these calculations, the central massive object is taken to be Sgr $A^{*}$ with mass $M_{\bullet}=4.31\times 10^{6}M_{\odot}$ and distance $D_{\rm OL}=8.33~\textrm{kpc}$, where $M_{\odot}$ is the mass of the sun. Further, the angles $\theta _{\infty}$ and s are expressed in units of micro-arcsecond ($\mu$as), and nano-arcsecond (nas) respectively. All the distances are measured with $GM_{\bullet}/c^{2}=1$ unit.}
\label{Table_01}       
%
%
\begin{tabular}{p{2.5cm}p{2.5cm}p{9.5cm}}
\hline\noalign{\smallskip}
\hline\noalign{\smallskip}
Observables                     & Schwarzschild  & ~~~~~~~~~~~~~~~~\KR Black  Hole       			                 \\
and                             &                & ~~~~~~~~~~~~~~~~~~~~~~~~~~~versus                                          \\
Parameters                      &                & ~~~~~~~~~~~~~~\RN Black Hole                                                \\
			         &		  &				                                                   \\
			         &             	  &  b=0.2~~ Q=0.2~~~~ b=0.5~~~ Q=0.5~~~ b=0.8~~ Q=0.8\\
\hline \noalign{\smallskip}
\hline \noalign{\smallskip}
$\theta _{\infty}$ ($\mu$as)     & 26.96         & 27.06 ~~~ 26.78  ~~~~ 27.21 ~~~ 25.78  ~~~~ 27.35 ~~~ 23.59 \\
s (nas)                          & 33.74         & 39.34 ~~~ 32.98 ~~~~ 50.18 ~~~ 30.57   ~~~~  65.25 ~~~ 36.71 \\
r (mag)                          & 6.82          & 6.76  ~~~~~~ 6.7 ~~~~~~~ 6.5 ~~~~~ 6.58  ~~~~~~  6.29 ~~~~~ 6.2   \\
$u_{\rm ph}(GM_{\bullet}/c^{2})$ & 5.20          & 5.22 ~~~~~ 5.16 ~~~~~ 5.24 ~~~~~ 4.97 ~~~~~~  5.27 ~~~~ 4.55  \\
$\bar{a}$                        & 1             & 1.02 ~~~~~ 1.01  ~~~~~~ 1.05 ~~~~~ 1.03  ~~~~~  1.08 ~~~~~ 1.12   \\
$\bar{b}$                        & -0.40         & -0.37 ~~~~ -0.41 ~~~~ -0.32 ~~~~ -0.68 ~~~~ -0.27 ~~~~ -0.98  \\
\noalign{\smallskip}
\hline\noalign{\smallskip}
\hline \noalign{\smallskip}
\end{tabular}
\end{table}
Given the premises for the current analysis, we will now present the exact results to leading order in $(b/r^{3})$, which will later be contrasted with the corresponding numerical estimates of the lensing observables obtained by inclusion of higher order terms. In this case the photon circular orbit can be obtained by solving the following simpler algebraic equation,
\begin{align}
r^{3}-3r^{2}-\frac{5b}{6}=0
\end{align}
Note that for $b=0$, i.e., if the \KR parameter identically vanishes one immediately arrives at $r_{\rm ph}=3$, the correct value for the photon radius in the Schwarzschild spacetime. The corresponding solution is given in terms of the \KR parameter $b$, to the leading order, as
\begin{align}
r_{\rm ph}&=\frac{\left(\sqrt{\left(\frac{5b}{3}\right)^{2}+\frac{10 b}{3}}+\frac{5b}{3}+2\right)^{1/3}}{\sqrt[3]{2}}+\frac{\sqrt[3]{2}}{\left(\sqrt{\left(\frac{5b}{3}\right)^{2}+\frac{10 b}{3}}+\frac{5b}{3}+2\right)^{1/3}}+1
\nonumber
\\
&\simeq 3+\left(\frac{5}{54}\right)b+\mathcal{O}(b^{2})
\end{align}
The theoretical variables associated with this model are the impact parameter associated with the photon circular orbit, i.e., $u_{\rm ph}=r_{\rm ph}/\sqrt{f(r_{\rm ph})}$ as well as the parameters $\bar{a}$ and $\bar{b}$ as mentioned in \ref{Eq_Lens_06b} and \ref{Eq_Lens_06c} respectively. For the evaluation of $\bar{b}$ as in \ref{Eq_Lens_06c}, one would require the regular part of the integral appearing in the expression for deflection angle, which to the leading order in the \KR parameter, yields, $b_{R}=0.9496+0.0528~ b$. Given all these ingredients, the analytical expressions for these variables in terms of the \KR parameter $b$, to the leading order, reads
\begin{align}
u_{\rm ph}&=\frac{r_{\rm ph}^{5/2}}{\sqrt{r_{\rm ph}^{3}-2r_{\rm ph}^{2}-\frac{b}{3}}}
\simeq 3\sqrt{3}\left[1+\frac{b}{54}+\mathcal{O}(b^{2})\right]
\\
\bar{a}&=\left[\frac{r_{\rm ph}^{3}-2r_{\rm ph}^{2}-\frac{b}{3}}{r_{\rm ph}^{3}-2r_{\rm ph}^{2}-br_{\rm ph}}\right]^{1/2}\left[1+\frac{5b}{3r_{\rm ph}^{3}}\right]^{-1/2}
\simeq 1+\left(\frac{19}{162}\right)b+\mathcal{O}(b^{2})
\\
\bar{b}&=-\pi +b_{\rm R}+\bar{a}\ln \left(\frac{2\left[r_{\rm ph}^{3}+\frac{5b}{3}\right]\left[\frac{r_{\rm ph}^{2}+\frac{b}{6}}{r_{\rm ph}^{2}+\frac{b}{2}}\right]^{2}}{r_{\rm ph}^{3}-2r_{\rm ph}^{2}-\frac{b}{3}} \right)
\simeq -0.4+0.19~b+\mathcal{O}(b^{2})
\end{align}
Note that these three variables are sufficient to determine the three observables $\theta _{\infty}$, $r$ and $s$, thanks to the relations presented in \ref{Eq_Lens_07a}, \ref{Eq_Lens_07b} and \ref{Eq_Lens_07c} respectively. The additional information necessary to have a numerical estimate are the mass and distance of the central object. The best such candidate corresponds to the supermassive black hole Sgr $A^{*}$ with mass $M_{\bullet}=4.31\times 10^{6}M_{\odot}$, where $M_{\odot}$ is the mass of the sun and $D_{\rm OL}=D_{\bullet}=8.33~\textrm{kpc}$. The numerical estimates of the observables $(\theta _{\infty},s,r)$ and the theoretical parameters $(u_{\rm ph},\bar{a},\bar{b})$ have been presented in \ref{Table_01} for various values of the \KR parameter $b$ along with the corresponding Schwarzschild as well as \RN values for comparison. One can immediately verify that the numerical values for $u_{\rm ph}$, $\bar{a}$ and $\bar{b}$, as presented in \ref{Table_01} are in excellent agreement with the 
corresponding ones obtained by keeping the leading order term in the \KR parameter alone.

As a final ingredient we present the explicit form of the deflection angle for the four-dimensional \KR black hole, which to the leading order in the \KR parameter becomes,
\begin{align}
\Theta \left(\theta\right)&=-2.192+0.0528~ b
\nonumber
\\
&+\left\lbrace \frac{r_{\rm ph}^{2}+\frac{b}{2}}{\left[1+\frac{5b}{3r_{\rm ph}^{3}}\right](r_{\rm ph}^{2}-br_{\rm ph}+\frac{5b}{6})}\right\rbrace^{1/2}\ln \left(\frac{2u_{\rm ph}\left[r_{\rm ph}^{3}+\frac{5b}{3}\right]\left[\frac{r_{\rm ph}^{2}+\frac{b}{6}}{r_{\rm ph}^{2}+\frac{b}{2}}\right]^{2}}{\left(r_{\rm ph}^{3}-2r_{\rm ph}^{2}-\frac{b}{3}\right)\left(\theta D_{\rm OL}-u_{\rm ph}\right)} \right)
\nonumber
\\
&\simeq -0.4-\ln \left(\frac{\theta D_{\rm OL}}{3\sqrt{3}}-1 \right)
+\left\lbrace 0.19 -\frac{19}{162}\ln \left(\frac{\theta D_{\rm OL}}{3\sqrt{3}}-1 \right) 
+\frac{1}{54}\left(\frac{1}{1-\frac{3\sqrt{3}}{\theta D_{\rm OL}}}\right)\right\rbrace~b
+\mathcal{O}(b^{2})
\end{align}
\begin{figure*}
\begin{center}

\includegraphics[height=2in, width=3in]{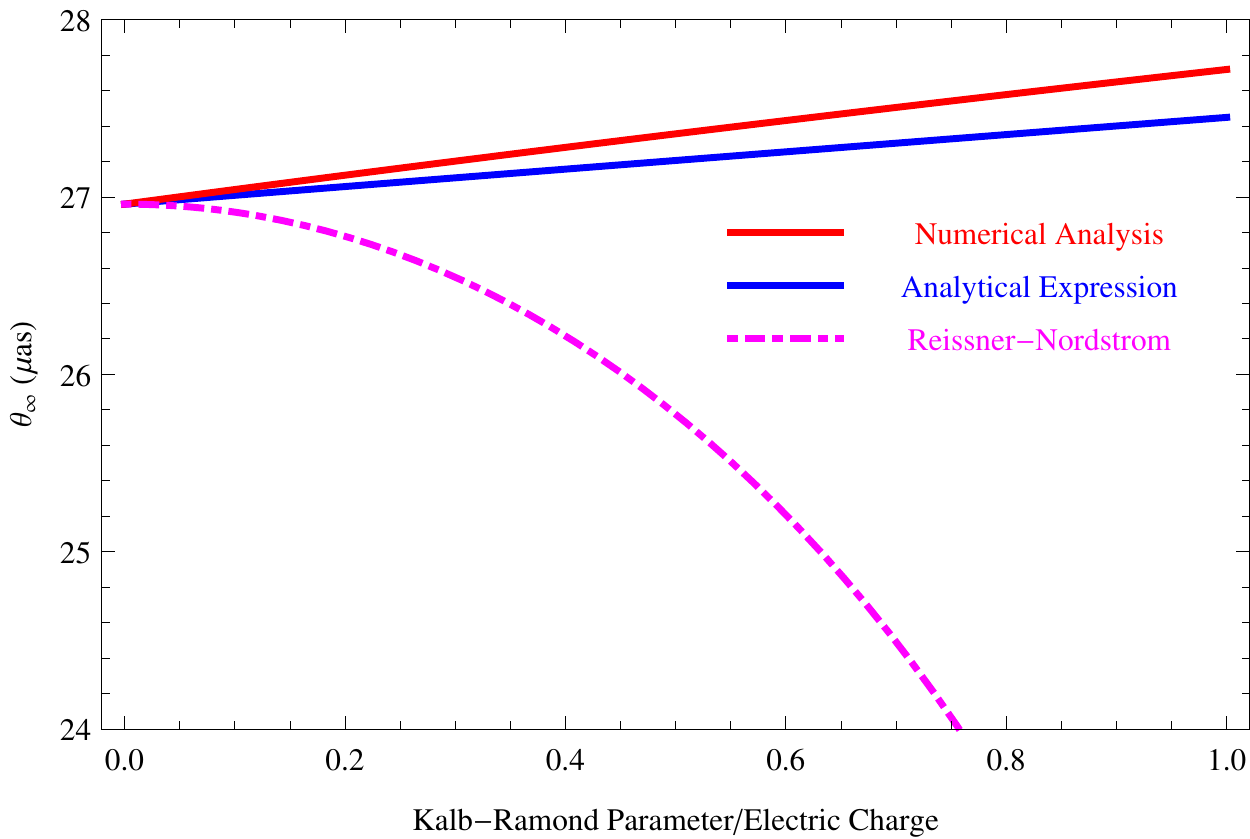}~~
\includegraphics[height=2in, width=3in]{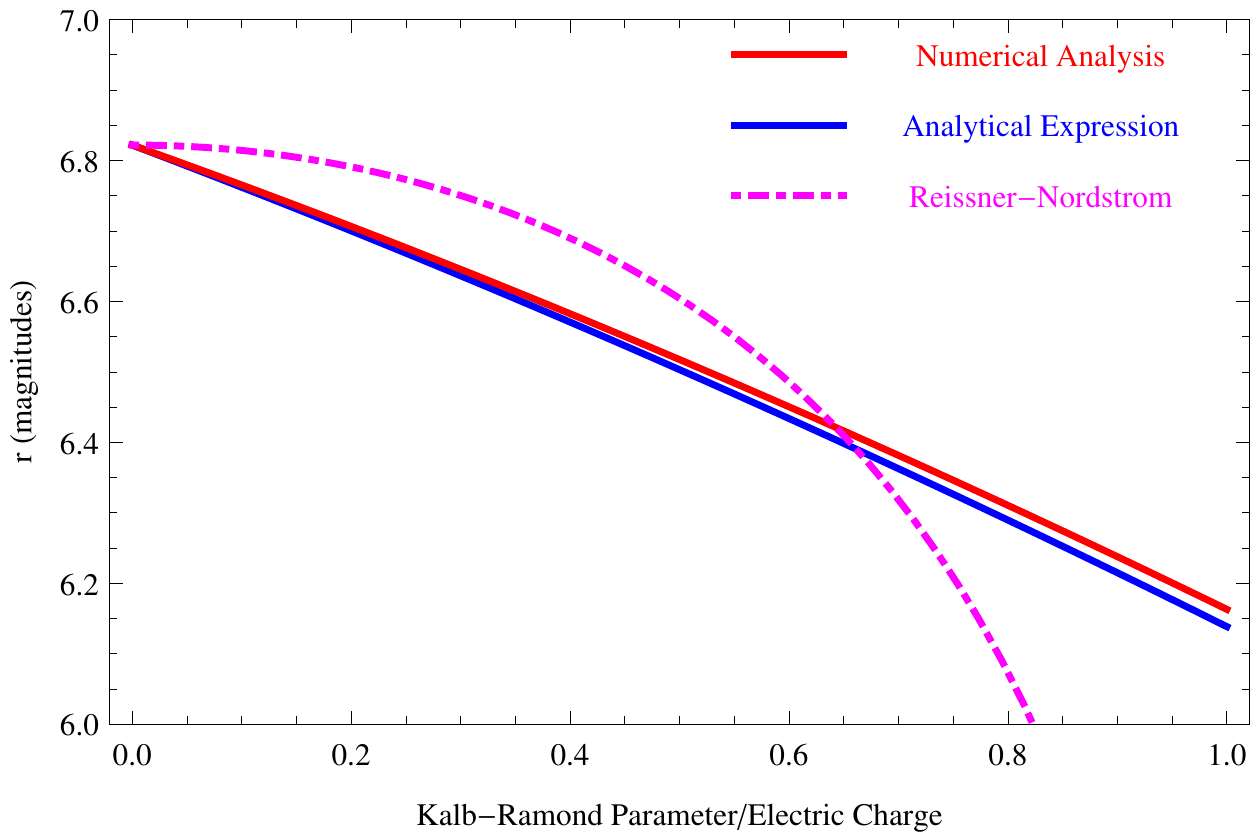}\\
\includegraphics[height=2in, width=3in]{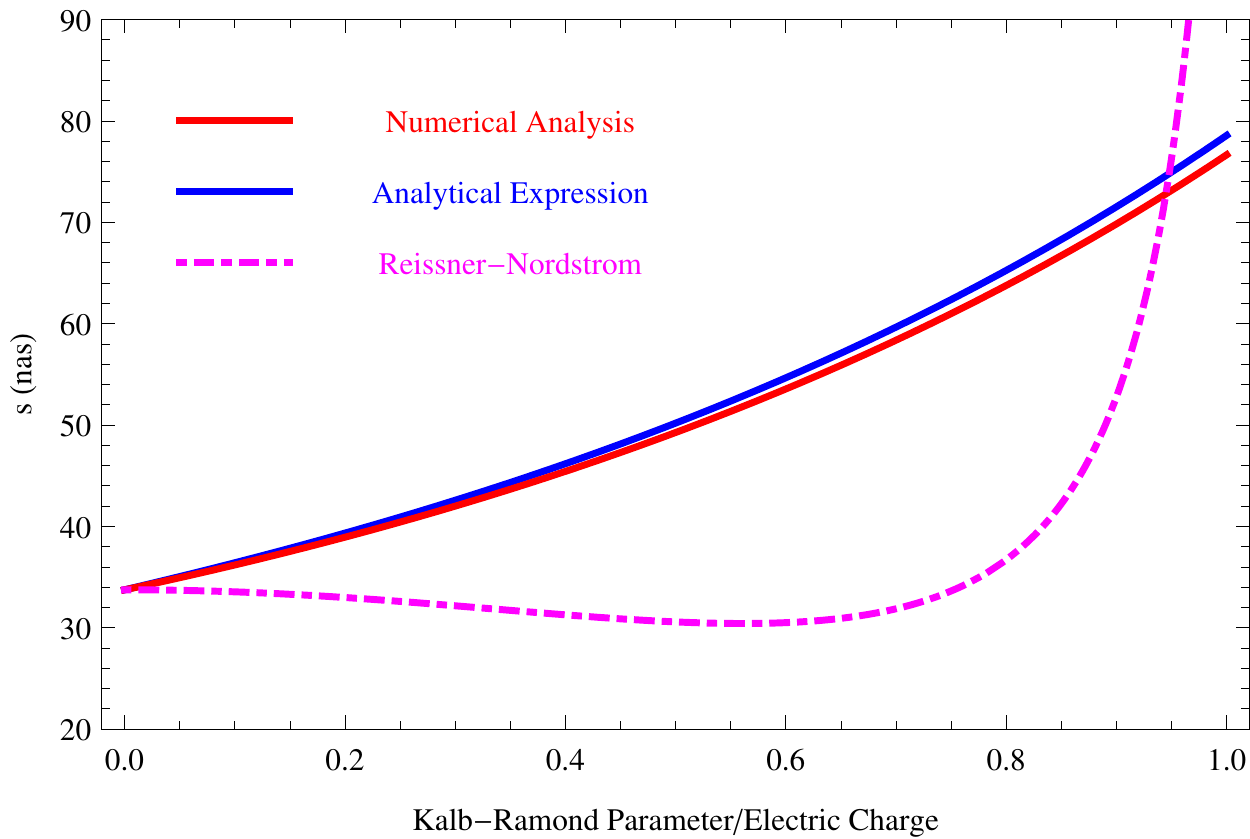}

\caption{Variation of the three observables --- (a) the deflection angle $\theta _{\infty}$ in micro-arcsecond, (b) the magnitude difference $r$ between the outermost and inner images and finally (c) the angular difference between the outermost and inner images $s$ in nano-arcsecond have been presented as the Kalb-Ramond parameter $b$, as well as the electric charge $Q$ increases. We have also presented both the approximate analytical and numerical estimates for these observables as well. For small \KR parameter they match very well, while for larger values of \KR parameter they exhibit minor deviations over the analytical result as expected. It is also evident that in four spacetime dimensions as the Kalb-Ramond parameter increases the corresponding deflection angle and angular separation increases, while the magnitude difference exhibits an opposite effect. Further the observables associated with \RN black hole show significant deviations from the \KR one. See text for more discussions.}
\label{Lensing_Fig_01}

\end{center}
\end{figure*}
So far, we have been dealing with our approximated metric elements, correct to $\mathcal{O}(b/r^{3})$. As promised, analytic expressions for the lensing observables as a function of the \KR parameter have been presented. However the strong lensing phenomenon originates from the region near the black hole horizon and one may wonder whether higher order terms in the metric elements can really be neglected. To see the effect of these terms explicitly, we have followed three steps --- (a) Solved the higher order algebraic equation for the photon circular orbit numerically; (b) The corresponding solution has been inserted in the expressions for $u_{\rm ph}$, $\bar{a}$ and $\bar{b}$ in terms of the metric elements as given in \ref{Eq_Lens_06a}, \ref{Eq_Lens_06b} and \ref{Eq_Lens_06c} respectively; (c) Then we have constructed the variation of the lensing observables with \KR parameter numerically and have plotted them in \ref{Lensing_Fig_01}. Surprisingly, from the variation of all the lensing 
observables with the \KR parameter as presented in \ref{Lensing_Fig_01} it turns out that the approximate analytic solution is in very good agreement with the numerical estimates. Further, when the \KR parameter is small, the analytical and numerical estimates are mostly identical, while deviations are more prominent as the \KR parameter becomes larger. This is expected, since the higher order terms become more dominant at large values of the \KR parameter.

Further, from both \ref{Table_01} and \ref{Lensing_Fig_01} one can understand the effect of \KR field on gravitational lensing both quantitatively and qualitatively. From \ref{Lensing_Fig_01} it is clear that the innermost deflection angle and angular separation increases monotonically as the \KR parameter is being increased, while the magnification parameter depicts an exactly opposite behaviour. Similar features are also present in \ref{Table_01}, but according to the numerical estimates, both the angular deflection and angular separation are larger compared to the Schwarzschild value. For example, when $b=0.5$, the deflection angle is almost one percent while the angular separation is $49$ percent larger. On the other hand, the magnification is almost five percent less compared to the Schwarzschild counterpart. A comparison with the \RN black hole reveals a similar story. From \ref{Table_01} it is clear that the value of $\theta _{\infty}$ for \KR black hole is much higher compared to 
the \RN black hole. For example, at $b=0.5$ (or, $Q=0.5$), the value of 
$\Theta _{\infty}$ allows a difference of more than five percent between 
\RN and \KR black hole. The same can also be verified from \ref{Lensing_Fig_01} as well.
\subsection{Kalb-Ramond field and extra dimensions}\label{KRfive}

Extra spatial dimensions are known to play crucial role in addressing the hierarchy problem. In some models of extra dimensions the matter fields are confined to the four-dimensional brane, while gravity extends over the higher dimensions. From stringy viewpoint the \KR field being a closed string mode like gravity, can also probe the extra spacetime dimensions. Hence in presence of both \KR field and extra dimensions it is important to consider the \KR field in the bulk spacetime. The natural question arising in this context is to address how the gravity looks like from a four-dimensional perspective in this scenario. This question has been answered recently in \cite{Chakraborty:2014fva}, where possible effects of higher dimensional \KR field in the context of brane world were studied. Further a static, spherically symmetric solution on the brane was also derived. In this section we consider strong gravitational lensing from a similar but slightly modified perspective. 

There will be two modifications in the effective gravitational field equations on the brane --- (a) effects of \KR field inherited from the bulk and (b) the effect of bulk Weyl tensor projected on the brane \cite{Chakraborty:2014fva}. The projected \KR field in presence of static and spherical symmetry is represented by a single function $h(r)$, which to the leading order behaves as $\sqrt{\tau/\kappa _{5}^{2}}(1/r^{2})$, where $\kappa _{5}$ is the five-dimensional gravitational constant and $\tau$ is the parameter characterizing the \KR field. With this solution for the \KR field the corresponding solution for the metric elements in the effective four dimensional spacetime on the brane becomes,
\begin{align}
f(r)&=1-\frac{2}{r}-\frac{a}{r^{2}}-\frac{2\tau}{r^{3}}-\frac{2a\tau}{r^{4}}+\mathcal{O}\left(\frac{\tau^{2}}{r^{5}}\right)
\\
g(r)&=1-\frac{2}{r}-\frac{a-\tau}{r^{2}}+\frac{a\tau}{3r^{4}}+\mathcal{O}\left(\frac{\tau^{2}}{r^{5}}\right)
\end{align}
where, $a=(3\alpha P_{0}/2)$ has been inherited from the bulk Weyl tensor, such that, $\alpha=(1/4\pi G_{N}\lambda _{T})$ with $\lambda _{T}$ being the brane tension and $P_{0}$ measures the effect of bulk Weyl tensor. Note that if the torsion parameter $\tau$ vanishes, the above solution reduces to the well known black hole solution on the brane derived in \cite{Dadhich:2000am}. 

\begin{figure*}
\begin{center}

\includegraphics[height=2in, width=3in]{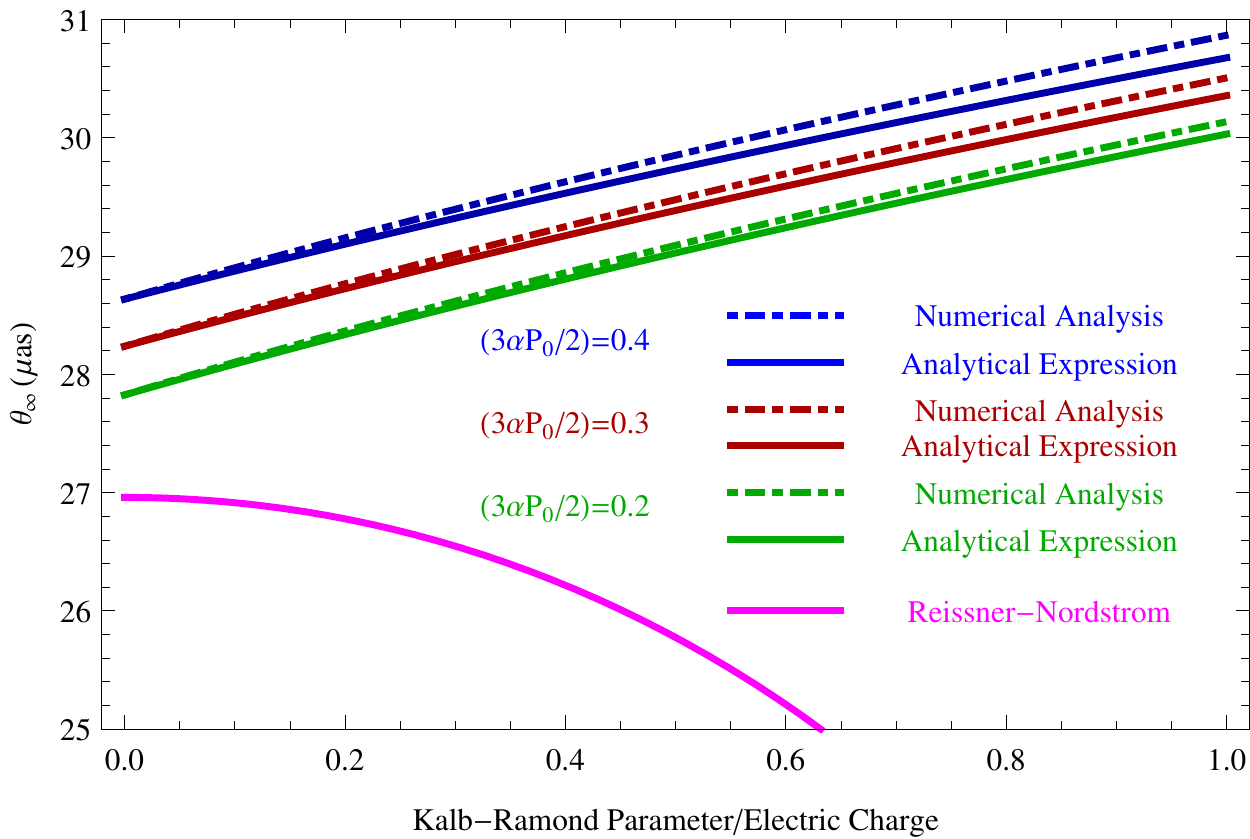}~~
\includegraphics[height=2in, width=3in]{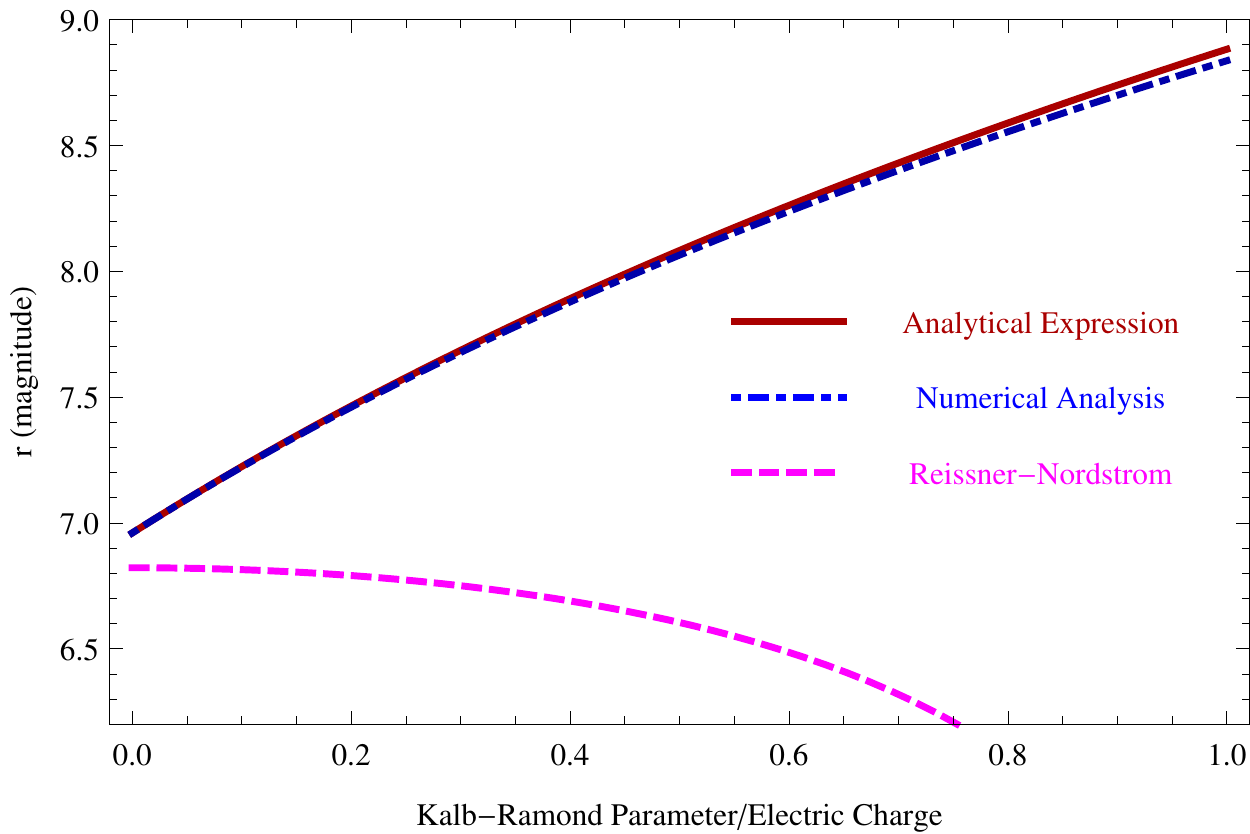}\\
\includegraphics[height=2in, width=3in]{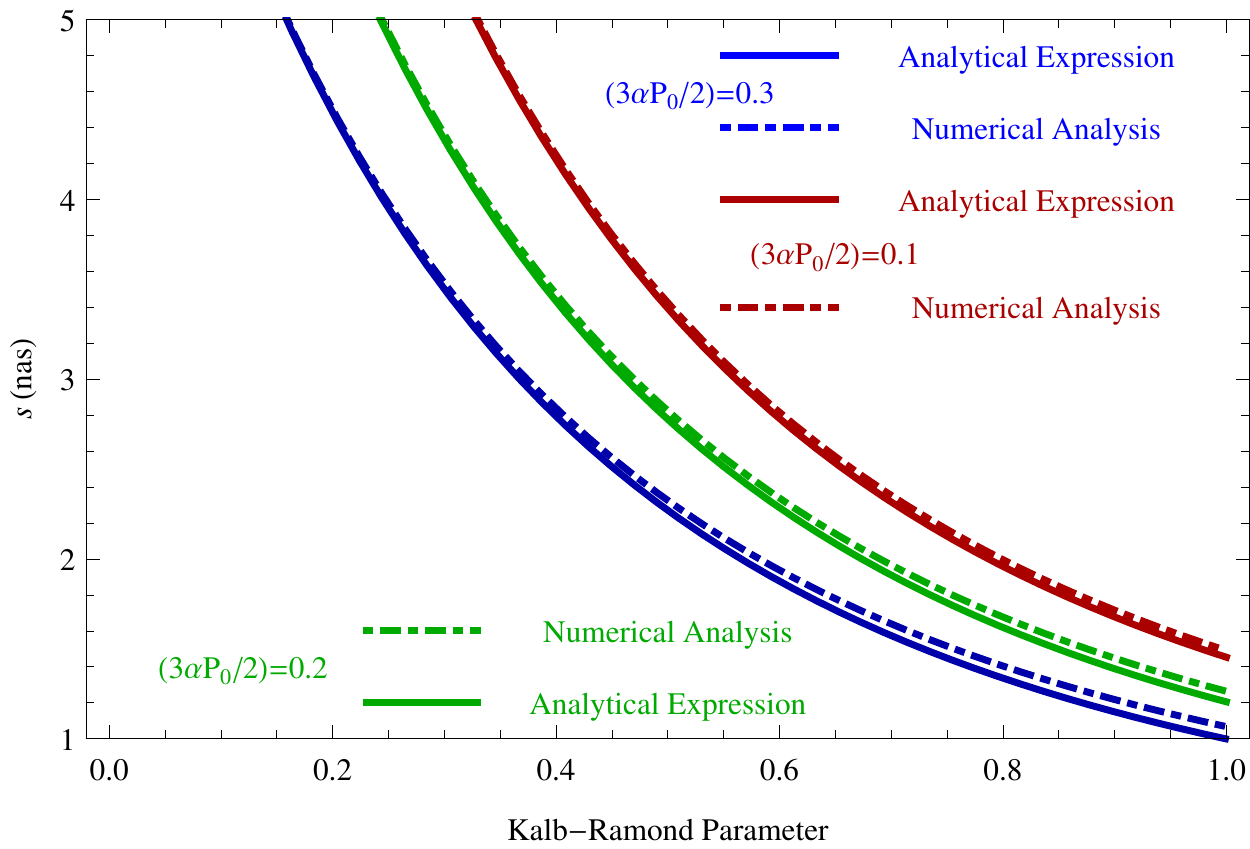}~~
\includegraphics[height=2in, width=3in]{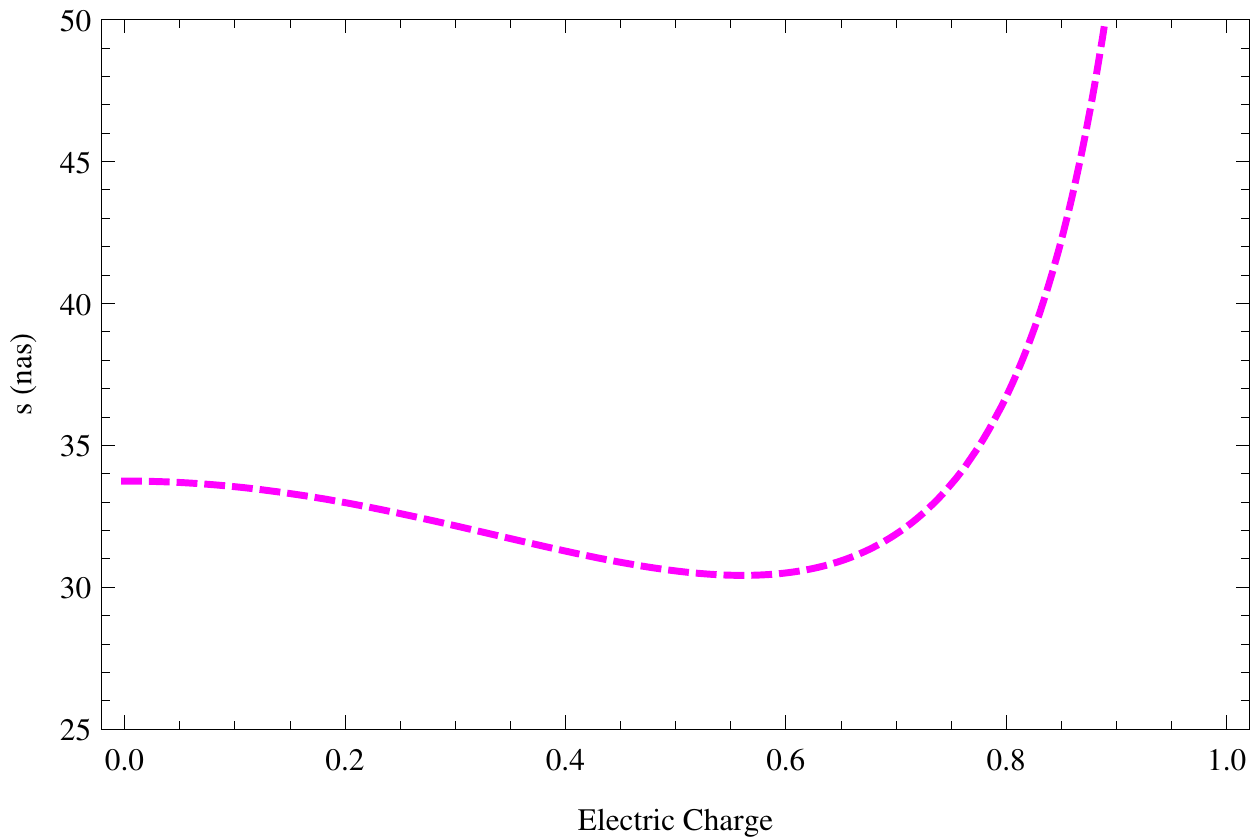}

\caption{Variation of the three observables --- (a) the deflection angle $\theta _{\infty}$ in micro-arcsecond, (b) the magnitude difference $r$ between the outermost and inner images and finally (c) the angular difference between the outermost and inner images $s$ in nano-arcsecond have been illustrated as the Kalb-Ramond parameter $\tau$ increases. We have also plotted the numerical estimates arising out of the higher order terms in all the three cases, which are in good agreement with the corresponding approximate analytical expressions. It is also clear that as the Kalb-Ramond parameter increases the corresponding deflection angle and magnitude difference also increases, while the angular separation exhibits an opposite effect. In addition, for comparison we have plotted the corresponding scenario for \RN black hole as well, which shows significant deviations from the \KR black hole. See text for more discussions.}\label{Lensing_Fig_02}

\end{center}
\end{figure*}

Having written down the metric, one needs to compute the photon circular orbit in this spacetime geometry. This can be obtained by using \ref{Eq_Lens_02} and leads to the following expression,
\begin{align}\label{Eq_Lens_Rev_02}
2=\frac{6}{r}+\frac{4a}{r^{2}}+\frac{10\tau}{r^{3}}+\frac{12a\tau}{r^{4}}+\mathcal{O}\left(\frac{\tau^{2}}{r^{5}}\right)
\end{align}
Given this quartic algebraic equation, the photon circular orbit will turn out to be highly complicated function of the bulk induced parameter $a$ and the torsion parameter $\tau$. The origin for such behavior can again be traced back to the fact that the metric elements presented above are expanded in powers of $(2/r)$, while we are interested in its nature near $r=3$, the location of photon circular orbit. This would force one to keep very high order terms in order to achieve an accurate result. Thus it will be very difficult to present the results in a closed form and hence in order to provide concise analytical expressions we have to make certain approximations. In this particular case, since both $a$ and $\tau$ are supposed to be small quantities, we can keep terms upto linear order in either of them. This suggests to keep terms upto $\mathcal{O}(a/r^{2})$ and $\mathcal{O}(\tau/r^{3})$ respectively in the metric elements. This brings down \ref{Eq_Lens_Rev_02} to a cubic equation,
\begin{align}
r^{3}-3r^{2}-2ar-5\tau=0
\end{align}
which can be solved to yield the radius of photon circular orbit to leading order in the dark pressure $a$ and \KR parameter $\tau$ as,
\begin{align}
r_{\rm ph}&=1+\frac{2^{1/3}(9+6a)}{3\left(54+54a+135\tau+\sqrt{\left(54+54a+135\tau\right)^{2}-4(9+6a)^{3}}\right)^{1/3}}
\nonumber
\\
&+\frac{\left(54+54a+135\tau+\sqrt{\left(54+54a+135\tau\right)^{2}-4(9+6a)^{3}}\right)^{1/3}}{3\times 2^{1/3}}
\nonumber
\\
&\simeq 3+\frac{2}{3}~a+\frac{5}{9}~\tau+\mathcal{O}(a^{2},\tau^{2})
\end{align}
The other quantities of importance are the three theoretical parameters for this model --- (a) the impact parameter at the photon radius, i.e., $u_{\rm ph}=r_{\rm ph}/\sqrt{f(r_{\rm ph})}$, (b) the two parameters $\bar{a}$ and $\bar{b}$ associated the gravitational deflection angle. These three parameters to the leading order in the dark pressure term $a$ and the \KR parameter $\tau$ read as follows, 
\begin{align}
u_{\rm ph}&=\frac{r_{\rm ph}^{5/2}}{\sqrt{r_{\rm ph}^{3}-2r_{\rm ph}^{2}-ar_{\rm ph}-2\tau}}
\simeq 3\sqrt{3}\left(1+\frac{1}{6}~a+\frac{1}{9}~\tau\right)+\mathcal{O}(a^{2},\tau^{2})
\\
\bar{a}&=r_{\rm ph}\sqrt{\frac{r_{\rm ph}^{3}-2r_{\rm ph}^{2}-ar_{\rm ph}-2\tau}{\left(r_{\rm ph}^{3}+2ar_{\rm ph}+10\tau\right)\left(r_{\rm ph}^{2}-2r_{\rm ph}+\tau-a\right)}}
\simeq 1-\frac{1}{9}~a-\frac{25}{54}~\tau+\mathcal{O}(a^{2},\tau^{2})
\\
\bar{b}&=-\pi +b_{R}+\bar{a}\ln \left[\frac{2\left(r_{\rm ph}^{3}+2ar_{\rm ph}+10\tau\right)\left(2r_{\rm ph}^{2}+ar_{\rm ph}+2\tau\right)^{2}}{\left(r_{\rm ph}^{3}-2r_{\rm ph}^{2}-ar_{\rm ph}-2\tau\right)\left(2r_{\rm ph}^{2}+2ar_{\rm ph}+6\tau\right)^{2}}\right]
\nonumber
\\
&\simeq -0.4 -2.01~a -0.73~\tau+\mathcal{O}(a^{2},\tau^{2})
\end{align}
where the numerical value of the regular part of the integral appearing in the deflection angle has been used, which reads $b_{R}=0.9496-1.594~a+0.318~\tau$. Then using \ref{Eq_Lens_07a}, \ref{Eq_Lens_07b} and \ref{Eq_Lens_07c} we can determine the three observables $\theta _{\infty}$, $s$ and $r$ respectively from the expressions of $u_{\rm ph}$, $\bar{a}$ and $\bar{b}$ derived above. As in the previous section here also we use sgr A$^{*}$ as the black hole candidate with its mass and distance being quoted in the earlier section. The numerical estimates of the observables as well as the theoretical parameters are presented in \ref{Table_02} for various values of the \KR parameter $\tau$ and the bulk inherited parameter $a$. Corresponding values of the parameters and observables are also presented for Schwarzschild spacetime for direct comparison. Moreover, the variation of the observables $\theta _{\infty}$, $s$ and $r$ with the \KR parameter $\tau$ are presented in \ref{Lensing_Fig_02}. 
\begin{table}
\caption{Numerical estimations of the lensing observables $(\theta _{\infty},s,r)$ and the parameters of the theory namely, $(u_{\rm ph},\bar{a},\bar{b})$ have been provided. Results for the Schwarzschild black hole as well as for the black hole in effective field theory for four possible choices of the higher dimensional Kalb-Ramond parameter $\tau$ and the bulk inherited parameter $a=3\alpha P_{0}/2$ are being presented. Numerical estimates of the lensing observables and model parameters for \RN black hole have also been presented. In all these calculations, the central black hole is assumed to be Sgr $A^{*}$ having mass $M_{\bullet}=4.31\times 10^{6}M_{\odot}$, which is $D_{\rm OL}=8.33~\textrm{kpc}$ away. In addition, the angle $\theta _{\infty}$ and $s$ are expressed in units of micro-arcsecond ($\mu$as) and nano-arcsecond (nas) respectively, for convenience. All the distances are presented in $GM_{\bullet}/c^{2}=1$ unit.}
\label{Table_02}       
%
%
\begin{tabular}{p{2.5cm}p{2.5cm}p{9.5cm}}
\hline\noalign{\smallskip}
\hline\noalign{\smallskip}
Observables                      & Schwarzschild &    Black hole with \KR field in effective theory       \\
and				  &               & ~~~~~~~~~~~~~~~~~~~~~~~~ versus                                 \\
Parameters                       &               & ~~~~~~~~~~ \RN black hole                                         \\
                                 &               & $\tau=0.2$ ~~ $Q=0.2$ ~~ $\tau=0.7$ ~~ $Q=0.7$~~ $\tau=0$ ~~~~ $\tau=0.9$   \\
                                 &               & $a=0$ ~~~~~~~~~~~~~~~~~~ $a=0.3$  ~~~~~~~~~~~~~~~ $a=0.4$ ~~ $a=0.4$    \\
\hline \noalign{\smallskip}
\hline\noalign{\smallskip}
$\theta _{\infty}$ ($\mu$as)     & 26.96         & 27.53 ~~~~~~ 26.78 ~~~~ 29.79 ~~~~~~ 23.84 ~~~~ 28.63  ~~~~ 30.50      \\
s (nas)                          & 33.74         & 8.75  ~~~~~~~ 32.98 ~~~~ 1.57  ~~~~~~~~ 35.89 ~~~~ 6.24 ~~~~~~ 0.95       \\
r (mag)                          & 6.82          & 7.39 ~~~~~~~ 6.7 ~~~~~~~  8.43  ~~~~~~~~ 6.34 ~~~~~~ 7.08  ~~~~~~ 8.72       \\
$u_{\rm ph}(GM_{\bullet}/c^{2})$ & 5.20          & 5.31 ~~~~~~~ 5.16 ~~~~~~ 5.74  ~~~~~~~ 4.62  ~~~~~~ 5.52 ~~~~~~ 5.89       \\
$\bar{a}$                        & 1             & 0.92 ~~~~~~~ 1.01 ~~~~~~ 0.81 ~~~~~~~ 1.09   ~~~~~~ 0.96 ~~~~~~ 0.78       \\
$\bar{b}$                        & -0.40         & -0.51 ~~~~~~ -0.41 ~~~~~ -1.12 ~~~~~~ -0.91 ~~~~ -1.17 ~~~~~~ -1.29      \\
\noalign{\smallskip}
\hline\noalign{\smallskip}
\hline\noalign{\smallskip}
\end{tabular}
\end{table}

The only remaining bit corresponds to the expression for the deflection angle in terms of the \KR parameter $\tau$ and the bulk charge $a$. This can be achieved by numerically computing the regular part of the integral appearing in the deflection angle, which . Thus finally the deflection angle, to leading order in the dark pressure $a$ and \KR parameter $\tau$ becomes,
\begin{align}
\Theta \left(\theta\right)&=-2.1920-1.594~a+0.318~\tau+r_{\rm ph}\sqrt{\frac{r_{\rm ph}^{3}-2r_{\rm ph}^{2}-ar_{\rm ph}-2\tau}{\left(r_{\rm ph}^{3}+2ar_{\rm ph}+10\tau\right)\left(r_{\rm ph}^{2}-2r_{\rm ph}+\tau-a\right)}}
\nonumber
\\
&\times \ln \left[\frac{\left(r_{\rm ph}^{3}+2ar_{\rm ph}+10\tau\right)\left(2r_{\rm ph}^{2}+ar_{\rm ph}+2\tau\right)^{2}}{\left(r_{\rm ph}^{3}-2r_{\rm ph}^{2}-ar_{\rm ph}-2\tau\right)\left(2r_{\rm ph}^{2}+2ar_{\rm ph}+6\tau\right)(\frac{\theta D_{\rm OL}}{u_{\rm ph}}-1)}\right]
\nonumber
\\
&\simeq -0.4 -\ln \left(\frac{\theta D_{\rm OL}}{3\sqrt{3}}-1\right)-\left[2.01-\frac{1}{9}\ln \left(\frac{\theta D_{\rm OL}}{3\sqrt{3}}-1\right)-\frac{1}{6}\left(\frac{1}{1-\frac{3\sqrt{3}}{\theta D_{\rm OL}}}\right) \right]~a
\nonumber
\\
&-\left\lbrace 0.73 -\frac{25}{54}\ln \left(\frac{\theta D_{\rm OL}}{3\sqrt{3}}-1\right)
-\frac{1}{9}\left(\frac{1}{1-\frac{3\sqrt{3}}{\theta D_{\rm OL}}}\right) \right\rbrace~\tau+\mathcal{O}(a^{2},\tau^{2})
\end{align}
At this stage it is worthwhile to discuss how accurate are the approximations made earlier, which were used to arrive at expressions for the lensing observables. To see this, we need to compute the lensing observables numerically, given the most general metric ansatz. For that purpose we have followed a route, identical to the one taken in the previous section, by first numerically solving for the photon circular orbit, then substituting the solution in the lensing observables and finally depicting the variation of them with the \KR parameter. The result of this numerical analysis has been presented in \ref{Lensing_Fig_02}. It is very clear that small deviations can be observed at larger values of the \KR parameter and for larger values of the bulk induced dark pressure, as expected. This suggests that the numerical estimates of the lensing observables, obtained by inclusion of higher order terms in the metric are very similar to their approximate analytical behaviour. This illustrates 
the accuracy of our analytical model.

Let us now discuss the effect of both the \KR parameter and the bulk effect on the lensing observables. From \ref{Table_02} it is clear that an increase in either of $\tau$ and $a$ increases the value of the deflection angle and magnification compared to the Schwarzschild one, while the angular separation decreases rapidly. The same behaviour is seen in \ref{Lensing_Fig_02} as well where it is clear that for a given $\tau$, the deflection angle increases as the bulk charge $a$ increase, while the opposite is seen in the angular separation. Note this this feature distinguishes from the previous model, where the angular separation increases while the magnification decreases. Thus gravitational lensing can act as a test bed for detection of extra dimensions, as the observables behave differently when extra dimensions are present. This might lead to interesting physics as the experimental observations more and more zooms in the near black hole region through future large telescopes. 
\subsection{Gravitational lensing in f(T) gravity}\label{fTgrav}

As the curvature invariants are replaced by Weitzenb\"{o}ck connection one obtains the teleparallel equivalent of general relativity and as a consequence the dynamics is governed by spacetime torsion \cite{Ferraro:2006jd,Ferraro:2008ey}. The invariant Lagrangian density under general coordinate transformation corresponds to $T$, which is second order in the torsion tensor. One can extend this framework by considering more general Lagrangian density $f(T)$, which has attracted much attention recently, thanks to its interesting cosmological predictions \cite{Daouda:2011rt,Capozziello:2001mq,Capozziello:2007ec,Myrzakulov:2012qp,Wanas:2012pu,
Capozziello:2010zz,Ferraro:2011us,Miao:2011ki}. For an interesting aspect of this in the context of black hole physics, see \cite{Lochan:2015bha}. 

Recently, various static and spherically symmetric solutions in the context of $f(T)$ gravity have been studied. In this work we will consider one such solution in four dimensions as the $f(T)$ gravity couples with the electromagnetic field. Since we are interested in asymptotically flat solutions, we will neglect the effect of positive cosmological constant and hence the metric elements become \cite{Capozziello:2012zj} 
\begin{align}\label{Eq_Torso_Rev}
f(r)&=1-\frac{2}{r}+\frac{\tilde{q}^{2}}{r^{2}}=g(r)
\end{align}
(for a derivation of this metric element from that given in \cite{Capozziello:2012zj} see \ref{App_02}). Interestingly, to lowest order, spherically symmetric solution in $f(T)$ gravity reduces to that of \RN black hole. However the formal similarity must not hide the conceptual difference between the two, in particular the origin of the charge term has completely different interpretation in the two cases. Here, $\tilde{q}$ is the torsion parameter, with the torsion tensor having a single independent component varying as $\tilde{q}^{2}/r^{4}$ (see \ref{App_02}). While in the case of \RN black hole the charge is of electromagnetic origin. We would again like to remind the reader that we are using units in which $GM=1$ and hence differs from the \RN solution by the numerical coefficient of the $1/r$ term only. Given the formal similarity with the \RN black hole one can compute all the lensing observables in an exact manner in this scenario, 
which we will not repeat here and can be found in \cite{Bozza:2002zj}.
\begin{table}
\caption{Lensing observables $(\theta _{\infty},s,r)$ and the parameters of the model, $(u_{\rm ph},\bar{a},\bar{b})$ have been estimated numerically. Results for both the Schwarzschild black hole and the black hole in $f(T)$ theory for three possible values of $\tilde{q}$ are presented. In all these calculations, the black hole is taken to be Sgr $A^{*}$ with mass $M_{\bullet}=4.31\times 10^{6}M_{\odot}$ and distance $D_{\rm OL}=8.33~\textrm{kpc}$ respectively. furthermore, the angle $\theta _{\infty}$ and s are expressed in units of micro-arcsecond ($\mu$as) and nano-arcsecond (nas) respectively. All the distances are measured with $GM_{\bullet}/c^{2}=1$ unit.}
\label{Table_03}       
%
%
\begin{tabular}{p{2.5cm}p{2.5cm}p{2.5cm}p{3.5cm}p{2.5cm}}
\hline\noalign{\smallskip}
\hline\noalign{\smallskip}
Observables                      & Schwarzschild &                 & Hole in $f(T)$ gravity &                 \\
and                              &               &                 &                        &                 \\
Parameters                       &               & $\tilde{q}=0.2$ & $\tilde{q}=0.5$        & $\tilde{q}=0.8$ \\
\hline \noalign{\smallskip}
\hline \noalign{\smallskip}
$\theta _{\infty}$ ($\mu$as)     & 26.96         & 26.78           & 25.78                  & 23.59           \\
s (nas)                          & 33.74         & 32.98           & 30.57                  & 36.71           \\
r (mag)                          & 6.82          & 6.7             & 6.5                    & 6.29            \\
$u_{\rm ph}(GM_{\bullet}/c^{2})$ & 5.20          & 5.16            & 4.97                   & 4.55            \\
$\bar{a}$                        & 1             & 1.01            &1.03                    & 1.12            \\
$\bar{b}$                        & -0.40         & -0.41           & -0.68                  & -0.98           \\
\noalign{\smallskip}
\hline\noalign{\smallskip}
\hline\noalign{\smallskip}
\end{tabular}
\end{table}
\begin{figure*}
\begin{center}

\includegraphics[height=2in, width=3in]{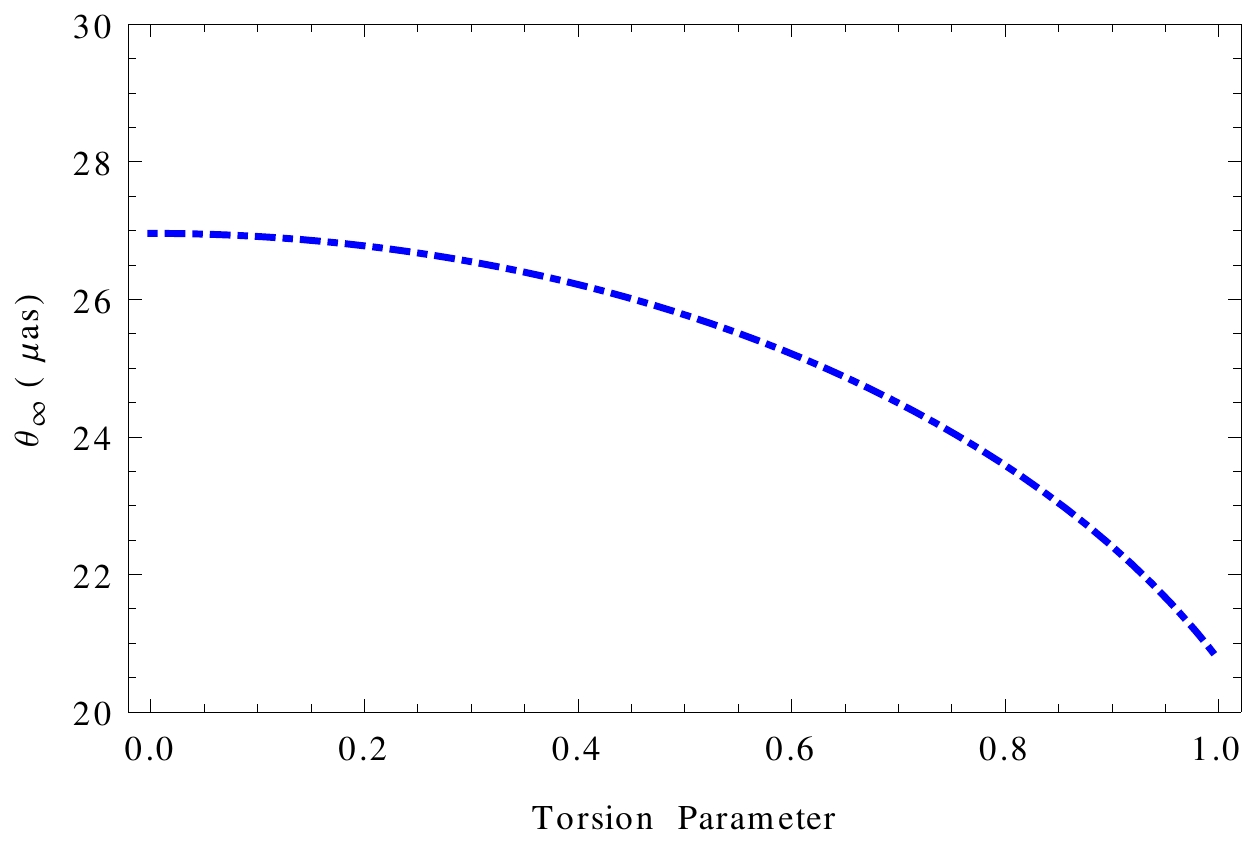}~~
\includegraphics[height=2in, width=3in]{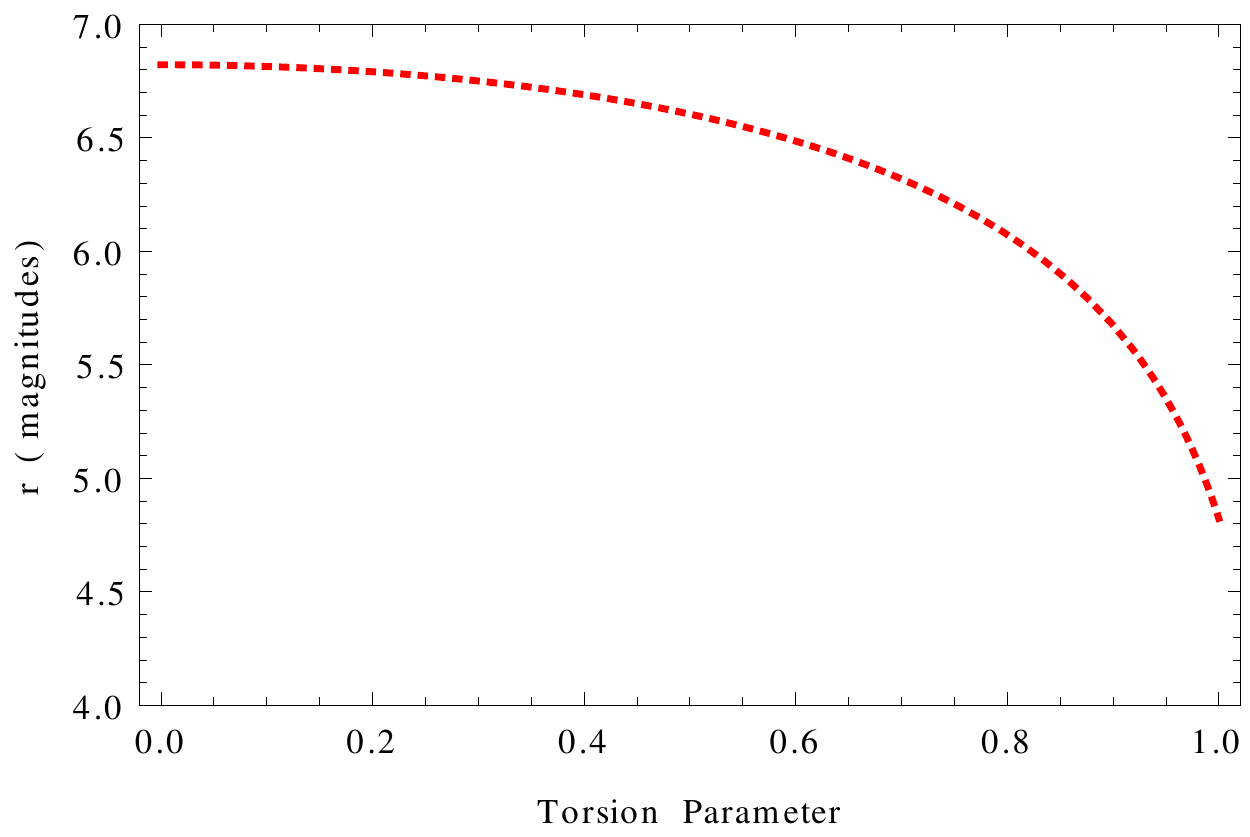}\\
\includegraphics[height=2in, width=3in]{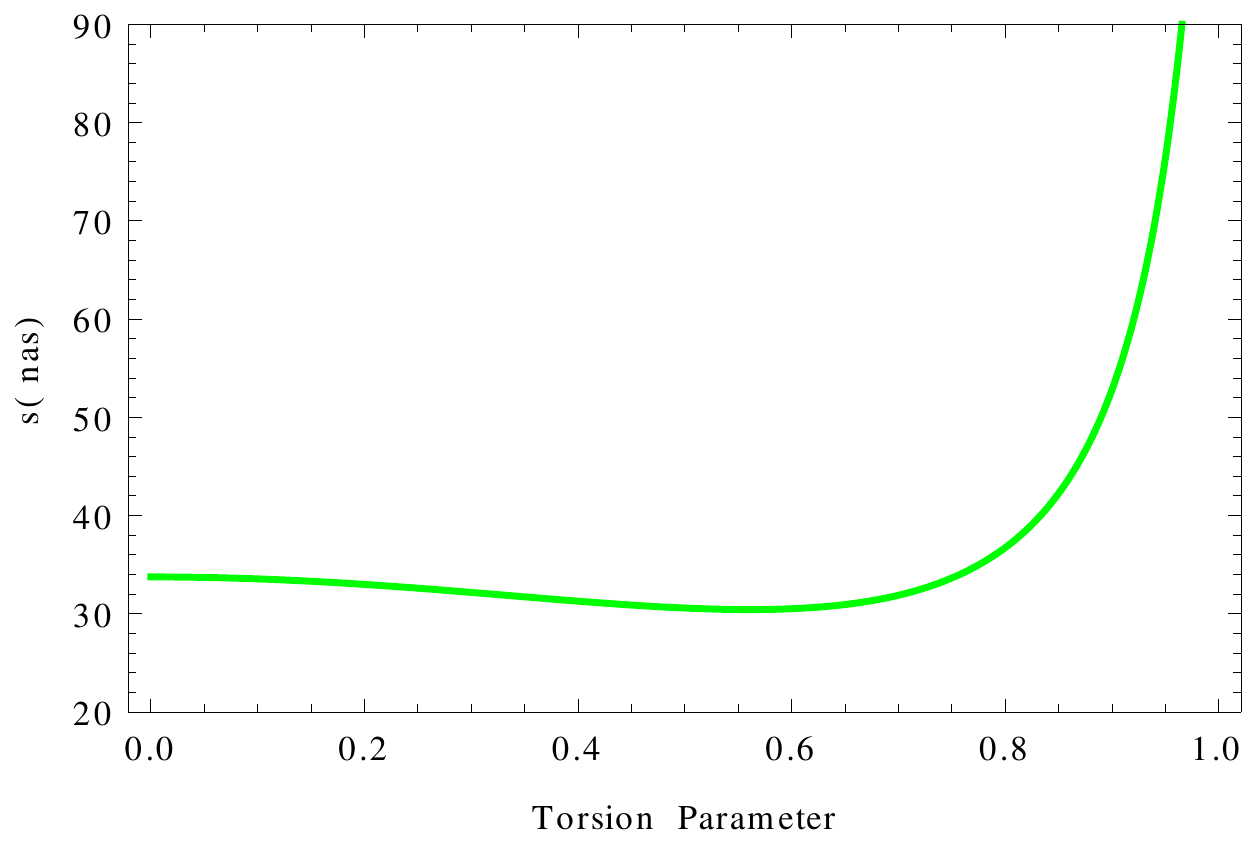}

\caption{Variation of the three observables, $\theta _{\infty}$ in micro-arcsecond, the magnitude difference $r$ and the angular separation between the outermost and inner images $s$ have been depicted with $\tilde{q}$. Hence in presence of $f(T)$ gravity in four spacetime dimensions with the increment of the torsion parameter, the corresponding deflection angle and magnitude difference decreases, while the angular separation initially decreases but then exhibits an opposite effect. See text for discussions.}\label{Lensing_Fig_03}

\end{center}
\end{figure*}

However for completeness we have presented numerical estimates for the lensing observables in \ref{Table_03} for three different values of the torsion parameter $\tilde{q}$ using the supermassive black hole Sgr $A^{*}$ in our galaxy as the gravitational lens. For convenience the corresponding values for $\tilde{q}=0$, i.e., for Schwarzschild spacetime are also depicted. Further variations of the three observables with $\tilde{q}$ are also illustrated in \ref{Lensing_Fig_03}. From \ref{Table_03} it is clear that the deflection angle and magnification decreases as $\tilde{q}$ increases, remaining always smaller compared to the Schwarzschild value. Note that this is exactly \emph{opposite} compared to the previous scenarios. On the other hand, the angular separation initially shows value small compared to Schwarzschild, but for large enough values of $\tilde{q}$ it increases. Similar features can be visualized through \ref{Lensing_Fig_03} as well. Again, with increase in resolution, future 
telescopes will be able to resolve the photon sphere surrounding the black hole. The particular shape one visualizes will depend on the strong field gravitational lensing and hence the observables discussed here can have important significance. 
\section{Concluding Remarks}

We have discussed the strong gravitational lensing in the context of both extra dimensions and \KR field. Specifically, we have discussed three situations --- (a) \KR field in four spacetime dimensions, (b) \KR field in higher dimensions and its effect on four dimensional brane inherited from the bulk and (c) possible effects of $f(T)$ gravity. In all the three situations we have provided explicit expressions for the three theoretical constructs $u_{\rm ph}$, $\bar{a}$ and $\bar{b}$ respectively. Using these analytical expressions we have estimated values for the three lensing observables $(\theta _{\infty},s,r)$ as well as have presented them graphically. In particular, analytical estimates of these three observables matches with the full numerical results, due to presence of higher order terms in the metric elements, very well. Further to understand possible discord and unity among the three models elaborated on in this work, we have presented collectively the behaviour of the observables in these three 
scenarios in \ref{Table_04}. 

It is clear that all the three scenarios are different in one aspect or another. The deflection angle $\theta _{\infty}$ increases for the \KR field in four and higher dimensions but decreases for $f(T)$ gravity. On the other hand, the magnification decreases for \KR field in four dimensions but increases in higher dimensions, while it increases for $f(T)$ gravity. The angular separation shows an opposite effect for \KR field in four and in higher dimensions, while changes from a decreasing to increasing mode in $f(T)$ gravity. Hence the three scenarios offer rich structures and one of the scenario can easily be differentiated from another. 

Having summarized the theoretical backdrops with numerical estimates of various observable parameters, let us now elaborate on possible observational signatures of the same. The most promising candidate in this direction seems to be the supermassive black hole at the galactic center, called Sgr A*, which we have used for our numerical estimates as well. Among others the star S2 and S6 orbiting Sgr A*, seems to be the best candidates to act as the source of the lensing and it has also been studied quiet methodically in \cite{Schodel:2002vg,Schodel:2003gy,Bozza:2004kq}. In particular one may encounter three possible scenarios --- (a) One can not distinguish the primary as well as the secondary images, (b) The primary image is well resolved but not the secondary image and (c) Both primary and secondary images are well resolved. It appears that most of the S stars orbiting the central supermassive black hole Sgr A*, the primary image would be resolvable but not the secondaries \cite{Bozza:2014ywa}. Note that the 
primary observable associated with all these observations is the angular separation between the lensed images and in our notation this corresponds to $s$. Surprisingly in our framework, the three black hole solutions have three different behaviors for the angular separation. For example, in the case of four-dimensional \KR field the angular separation must be larger compared to the Schwarzschild value while for the higher dimensional \KR field angular separation decreases. On the other hand, for $f(T)$ gravity, the angular separation initially decreases with the torsion parameter but ultimately it starts increasing. Thus in the future if one can arrive at an angular separation which is larger than the Schwarzschild value, then possible existence of higher dimensional \KR field can be ruled out. However as evident from the numerical calculations, the observed angular displacement of the lensed images turns out to be in the range $\sim 20-30$ micro arcsecond with a magnification of $6-8$ which are 
well outside the resolutions of the present day astronomical instruments. However the future experiments involving very long baseline interferometry are designed to have an accuracy of $10-100$ micro arcsecond along with Milli arcsecond angular resolution imaging 
\cite{Bozza:2012by,Gillessen:2009ht,Broderick:2009ph,Gillessen:2008qv,Ghez:2005dd,Weinberg:2004nj}. 
In particular it can be demonstrated using the techniques developed in \cite{Bozza:2004kq,Bozza:2005sc} that the star S6 orbiting the Sgr A* supermassive black hole will produce a gravitationally lensed image at an angular separation of 30 micro arcsecond by 2062 \cite{Bozza:2012by}.

Thus in the near future, with either event horizon or square kilometer array operational one can directly measure the photon sphere of the sgr $A^{*}$ supermassive black hole and hence the observables associated with strong gravitational lensing can be experimentally determined. This will lead to possible constraints on the \KR field parameter as well as spacetime torsion due to strong gravitational lensing. In particular with sufficiently better data one might vote in favour of one particular scenario compared to others. For example, if one observes that the magnification of the image increases compared to the Schwarzschild value, then that will definitely bring forward the scenario of \KR field in higher spacetime dimensions. On the other hand, if the magnification decreases but the deflection angle becomes larger than the Schwarzschild value, then \KR field in four spacetime dimensions would a viable candidate. Hence the study of strong gravitational lensing has the potential to address 
some of the 
fundamental questions of the universe, e.g., are there extra spacetime dimensions, why spacetime torsion (or, the \KR field) is missing in our nature, etc.? These would be worthwhile to study in the future. 
\begin{table}
\caption{A comparison between the three scenarios}
\label{Table_04}       
%
%
\begin{tabular}{p{4.4cm}p{3.5cm}p{3.5cm}p{3.0cm}}
\hline\noalign{\smallskip}
\hline\noalign{\smallskip}
Scenario                                 & Behaviour                        & Behaviour            & Behaviour              \\
under                                    &      of                          &      of              &      of                \\
consideration                            & deflection ($\theta _{\infty}$)  & separation ($s$)     & magnification  ($r$)   \\
\hline \noalign{\smallskip}
Kalb-Ramond field in four dimensions     & Increases                        & Increases            & Decreases              \\
Kalb-Ramond field in higher dimensions   & Increases                        & Decreases            & Increases              \\
$f(T)$ gravity                           & Decreases                        & Initially decreases  & Decreases              \\
                                         &                                  & then increases       &                        \\
\hline\noalign{\smallskip}
\hline\noalign{\smallskip}
\end{tabular}
\end{table}
\section*{Acknowledgements}

The authors sincerely thank the anonymous referee for helpful comments. Research of S.C. is supported by the SERB-NPDF grant (PDF/2016/001589) from SERB, Government of India.

\appendix
\labelformat{section}{Appendix #1} 
\labelformat{subsection}{Appendix #1}
\section{Appendix: Derivation of the deflection angle}\label{App_01}

We will work with the metric ansatz as presented in \ref{Eq_Lens_01}, which represents the most general static and spherically symmetric spacetime. Note that the original calculation for a general static and spherically symmetric spacetime has been performed by Bozza in \cite{Bozza:2002zj} which placed strong field gravitational lensing on a firm footing. The results presented in this appendix is by and large following the lines of \cite{Bozza:2002zj}, however we will write it in an independent manner with slight modifications over \cite{Bozza:2002zj}. For the motion of a photon (or a null ray) in this spacetime, the geodesic equations correspond to,
\begin{align}
\left(\frac{dr}{d\lambda}\right)^{2}&=\frac{g(r)}{f(r)}\left[E^{2}-f(r)\frac{L^{2}}{r^{2}}\right]
\\
\frac{d\phi}{d\lambda}&=\frac{L}{r^{2}}
\end{align}
where $E$ and $L$ are the conserved energy and the conserved angular momentum respectively. From these two equations one can eliminate the unknown parameter $\lambda$ and hence obtain the orbit equation for the null ray, which is given in \ref{Eq_Lens_03}. A slightly modified version of the orbit equation can be obtained by introducing the impact parameter $b=L/E$ as well as identifying it with a radius $r_{0}$, such that $b=r_{0}/\sqrt{f(r_{0})}$. Integrating this modified orbit equation one obtains the deflection angle as given by \ref{Eq_Lens_05}. The deflection angle has a divergent part given by,
\begin{equation}
F(x,r_{0})=\left[f(r_{0})-\frac{r_{0}^{2}}{r^{2}}f(r)\right]^{-1/2}
\end{equation}
where $x=x(r)$, is an arbitrary function of the radial coordinate $r$, such that $x(r_{0})=0$. Note that the term within inverse square root vanishes as $r\rightarrow r_{0}$ and hence $F(x=0,r_{0})$ diverges. Let us now try to expand the term inside square root upto quadratic order in $x$, which captures the difference $r-r_{0}$. The first term in the Taylor expansion corresponds to,
\begin{align}
\alpha &=\dfrac{d}{dx}\left[f(r_{0})-\frac{r_{0}^{2}}{r^{2}}f(r)\right]
=r_{0}^{2}\left[-\frac{f'(r)}{r^{2}}+\frac{2f(r)}{r^{3}}\right]\frac{dr}{dx}
\end{align}
Note that when $r=r_{\rm ph}$, the term vanishes. While the second term yields 
\begin{align}
\beta &=\frac{1}{2}\dfrac{d^{2}}{dx^{2}}\left[f(r_{0})-\frac{r_{0}^{2}}{r^{2}}f(r)\right]
=\frac{1}{2}\frac{dr}{dx}\dfrac{d\alpha}{dr} 
\nonumber
\\
&=\frac{r_{0}^{2}}{2}\left[\left(\frac{dr}{dx}\right)^{2}\left\lbrace -\frac{f''(r)}{r^{2}}+\frac{4f'(r)}{r^{3}}-\frac{6f(r)}{r^{4}} \right\rbrace +\frac{dr}{dx}\left\lbrace -\frac{f'(r)}{r^{2}}+\frac{2f(r)}{r^{3}} \right\rbrace \dfrac{d}{dr}\frac{dr}{dx}\right]
\end{align}
On $r=r_{\rm ph}$, where $r_{\rm ph}f'(r_{\rm ph})=2f(r_{\rm ph})$ the above turns out to be,
\begin{align}
\beta _{\rm ph}&=\frac{r_{0}^{2}}{2}\left(\frac{dr}{dx}\right)_{r_{\rm ph}}^{2}\left\lbrace -\frac{f''(r_{\rm ph})}{r_{\rm ph}^{2}}+\frac{2f(r_{\rm ph})}{r_{\rm ph}^{4}} \right\rbrace
\end{align}
Using the above trick one can separate out the integral in \ref{Eq_Lens_05} to a divergent and a regular part which yields (see also \cite{Bozza:2002zj}), 
\begin{align}
I(r_{0})&=\int _{0}^{x_{\infty}}dx \left[\left\lbrace R(x,r_{0})F(x,r_{0})-R(0,r_{\rm ph})F_{0}(x,r_{0})\right\rbrace +R(0,r_{\rm ph})F_{0}(x,r_{0}) \right] 
\\
&=I_{R}(r_{0})+I_{D}(r_{0});\qquad F_{0}(x,r_{0})=\frac{1}{\sqrt{\alpha x+\beta x^{2}}}
\end{align}
where, $r_{\rm ph}$ is the radius of photon circular orbit and is a solution of \ref{Eq_Lens_02}. Given this particular decomposition, the divergent integral $I_{D}(r_{0})$ can be solved explicitly, leading to,
\begin{align}
I_{D}(r_{0})&=\frac{R(0,r_{\rm ph})}{\sqrt{\beta}}\ln \left[ \frac{\alpha +2\beta +2\sqrt{\alpha \beta +\beta ^{2}}}{\alpha}\right]
\nonumber
\\
&=\frac{2R(0,r_{\rm ph})}{\sqrt{\beta}}\ln \left[\frac{\sqrt{\beta}+\sqrt{\alpha +\beta}}{\sqrt{\alpha}}\right]
\end{align}
Expanding the above expression around $r_{\rm ph}$ we obtain,
\begin{align}
I_{D}(r_{0})&=-\frac{R(0,r_{\rm ph})}{\sqrt{\beta _{\rm ph}}}\ln \left[\alpha _{\rm ph}+\alpha _{\rm ph}'r_{\rm ph} \left(\frac{r_{0}}{r_{\rm ph}}-1\right)\right]+\frac{R(0,r_{\rm ph})}{\sqrt{\beta _{\rm ph}}}\ln \left(4\beta _{\rm ph} \right)+\mathcal{O}\left(r_{0}-r_{\rm ph}\right)
\nonumber
\\
&=-\frac{R(0,r_{\rm ph})}{\sqrt{\beta _{\rm ph}}}\ln \left(\frac{r_{0}}{r_{\rm ph}}-1\right)
-\frac{R(0,r_{\rm ph})}{\sqrt{\beta _{\rm ph}}}\ln \left(\frac{\alpha _{\rm ph}'r_{\rm ph}}{4\beta _{\rm ph}}\right)+\mathcal{O}\left(r_{0}-r_{\rm ph}\right)
\end{align}
where we have used the result that $r_{\rm ph}=0$. From the above discussion it is evident that $\alpha _{\rm ph}'=(2\beta _{\rm ph})/(dr/dx)_{\rm ph}$. Thus finally we obtain, 
\begin{align}
I_{D}(r_{0})&=-a\ln \left(\frac{r_{0}}{r_{\rm ph}}-1\right)+b_{D}+\mathcal{O}\left(r_{0}-r_{\rm ph}\right)
\end{align}
where (following \cite{Bozza:2002zj})
\begin{align}
a=\frac{R(0,r_{\rm ph})}{\sqrt{\beta _{\rm ph}}};\qquad 
b_{D}=-\frac{R(0,r_{\rm ph})}{\sqrt{\beta _{\rm ph}}}\ln \left(\frac{r_{\rm ph}}{2\left(\frac{dr}{dx}\right)}\right)
\end{align}
Let us now introduce another new variable $u$ equivalent to the impact parameter as,
\begin{align}
u=\frac{r}{\sqrt{f(r)}}
\end{align}
such that when Taylor expanded the co-efficients of the first two term at $r=r_{\rm ph}$ becomes 
\begin{align}
\left(\frac{du}{dr}\right)_{r_{\rm ph}}=\left[\frac{1}{\sqrt{f}}-\frac{1}{2}\frac{rf'(r)}{f^{3/2}(r)}\right]_{r_{\rm ph}}=0
\end{align}
and
\begin{align}
\frac{1}{2}\left(\frac{d^{2}u}{dr^{2}}\right)_{r_{\rm ph}}=\frac{1}{2}\left[-\frac{f'(r)}{f^{3/2}(r)}-\frac{rf''(r)}{2f^{3/2}(r)}+\frac{3}{4}\frac{rf'^{2}(r)}{f^{5/2}(r)}\right]_{r_{\rm ph}}
=\frac{1}{4r_{\rm ph}f^{3/2}(r_{\rm ph})}\left[2f(r_{\rm ph})-r_{\rm ph}^{2}f''(r_{\rm ph})\right]
\end{align}
Thus,
\begin{align}
u-u_{\rm ph}=c\left(r-r_{\rm ph}\right)^{2}
\end{align}
where,
\begin{align}
c=\frac{1}{4r_{\rm ph}f^{3/2}(r_{\rm ph})}\left[2f(r_{\rm ph})-r_{\rm ph}^{2}f''(r_{\rm ph})\right]=\frac{r_{\rm ph}^{3}\beta _{\rm ph}}{2r_{0}^{2}f^{3/2}(r_{\rm ph})}\left(\frac{dr}{dx}\right)_{r_{\rm ph}}^{-2}
\end{align}
Hence the divergent integral turns out to be (see \cite{Bozza:2002zj} for more details),
\begin{align}
I_{D}(r_{0})&=-a\ln \left(\frac{\sqrt{u_{\rm ph}}}{r_{\rm ph}\sqrt{c}}\sqrt{\frac{u_{0}}{u_{\rm ph}}-1}\right)+b_{D}+\mathcal{O}\left(r_{0}-r_{\rm ph}\right)
\nonumber
\\
&=-\frac{a}{2}\ln \left(\frac{u_{0}}{u_{\rm ph}}-1\right)-a\ln \left(\frac{r_{\rm ph}}{2\left(\frac{dr}{dx}\right)}\right)-a\ln \left(\frac{\sqrt{u_{\rm ph}}}{r_{\rm ph}\sqrt{c}}\right) +\mathcal{O}\left(r_{0}-r_{\rm ph}\right)
\nonumber
\\
&=-\frac{a}{2}\ln \left(\frac{u_{0}}{u_{\rm ph}}-1\right)-\frac{a}{2}\ln \left(\frac{u_{\rm ph}}{4c\left(\frac{dr}{dx}\right)^{2}}\right)
\end{align}
Substituting for $c$, we immediately arrived at,
\begin{align}
I_{D}(r_{0})&=-\frac{a}{2}\ln \left(\frac{u_{0}}{u_{\rm ph}}-1\right)-\frac{a}{2}\ln \left[\frac{r_{\rm ph}}{4\sqrt{f(r_{\rm ph})}}\frac{1}{\left(\frac{dr}{dx}\right)^{2}}\left(\frac{r_{\rm ph}^{3}\beta _{\rm ph}}{2r_{0}^{2}f^{3/2}(r_{\rm ph})}\right)^{-1}\left(\frac{dr}{dx}\right)_{r_{\rm ph}}^{2}\right]
\nonumber
\\
&=-\frac{a}{2}\ln \left(\frac{u_{0}}{u_{\rm ph}}-1\right)+\frac{a}{2}\ln \left(\frac{2r_{\rm ph}^{2}}{r_{0}^{2}}\frac{\beta _{\rm ph}}{f(r_{\rm ph})} \right)
\end{align}
Now if $\theta$ is the angle between the image and the lens, one readily obtains the following result for deflection angle in terms of $\theta$ as in \ref{Lens_Main}. The objects present in the deflection angle has the following expressions
\begin{align}
\bar{a}=\frac{R(0,r_{\rm ph})}{2\sqrt{\beta _{\rm ph}}};\qquad \bar{b}=-\pi +b_{R}+\bar{a}\ln \left(\frac{2\beta _{\rm ph}}{f(r_{\rm ph})} \right)
\end{align}
where the quantity $b_{R}$ corresponds to,
\begin{align}
b_{R}=I_{R}(r_{\rm ph})=\int _{x_{\rm ph}}^{x_{\infty}}dx \left\lbrace R(x,r_{\rm ph})F(x,r_{\rm ph})-R(0,r_{\rm ph})F_{0}(x,r_{\rm ph})\right\rbrace
\end{align}
These are used in order to obtain \ref{Eq_Lens_06a}, \ref{Eq_Lens_06b} and \ref{Eq_Lens_06c}.
\section{Appendix: Derivation of the metric element in \ref{Eq_Torso_Rev}}\label{App_02}

The spherically symmetric spacetime geometry as in \ref{Eq_Lens_01}, in the context of $f(T)$ gravity in $D$ spacetime dimensions has been derived in \cite{Capozziello:2012zj}. From which the metric elements as well as the torsion field have the following solutions,
\begin{align}
T(r)&=\frac{-1\pm \sqrt{1-24\alpha \Lambda -6\alpha Q^{2}r^{4-2D}}}{6\alpha}
\end{align}
\begin{align}
g(r)&=\left(\frac{2}{3}\pm\frac{1}{3}\sqrt{1-24\alpha \Lambda -6\alpha Q^{2}r^{4-2D}}\right) ^{-2}r^{3-D}
\Bigg( \frac{1}{\left( D-2\right) }
\nonumber
\\
&\times \Bigg\{\frac{1}{54\alpha} \Big[-\frac{18\alpha Q^{2}r^{3-D}}{3-D}-\frac{\left( 1+72\alpha \Lambda \right) r^{D-1}}{D-1}\Big]
\nonumber
\\
&\pm \frac{\sqrt{r^{4D}\left( 1-24\alpha \Lambda-6\alpha Q^{2}r^{4-2D}\right) }}{54\alpha}\Big[ \frac{6\alpha Q^{2}r^{3-3D}}{2D-5}-\frac{\left(-1+24\alpha \Lambda \right) r^{-1-D}}{D-1}\Big]
\nonumber
\\
&\mp \frac{\left( D-2\right) ^{2}\left( -1+24\alpha \Lambda \right) Q^{2}r^{3+D} \sqrt{1+\frac{6\alpha Q^{2}r^{4-2D}}{-1+24\alpha \Lambda }}\ _2F_1\left(\frac{D-3}{2\left( D-2\right) },\frac{1}{2},\frac{3D-7}{2\left( D-2\right)};\frac{6\alpha Q^{2}r^{4-2D}}{1-24\alpha \Lambda }\right)}{3\left(D-3\right) 
\left( 2D-5\right) \left( D-1\right) \sqrt{r^{4D}\left(1-24\alpha \Lambda -6\alpha Q^{2}r^{4-2D}\right)}}\Bigg\}
\nonumber\\
&+\textrm{constants}\Bigg) 
\end{align}
\begin{align}
f(r)&=\frac{1}{r^{D-3}}\Bigg(\frac{1}{\left( D-2\right)}\Bigg\{ \frac{1}{54\alpha} 
\Big[-\frac{18\alpha Q^{2}r^{3-D}}{3-D}-\frac{\left( 1+72\alpha \Lambda\right) r^{D-1}}{D-1}\Big]
\nonumber
\\
&\pm \frac{\sqrt{r^{4D}\left( 1-24\alpha \Lambda-6\alpha Q^{2}r^{4-2D}\right) }}{54\alpha}\Big[\frac{6\alpha
Q^{2}r^{3-3D}}{2D-5}-\frac{\left(-1+24\alpha \Lambda \right) r^{-1-D}}{D-1}\Big]
\nonumber
\\
& \mp \frac{\left(D-2\right)^{2}\left(-1+24\alpha \Lambda \right) Q^{2}r^{3+D}
\sqrt{1+\frac{6\alpha Q^{2}r^{4-2D}}{-1+24\alpha \Lambda }}\ _2F_1\left(\frac{D-3}{2\left( D-2\right) },\frac{1}{2},\frac{3D-7}{2\left( D-2\right)};\frac{6\alpha Q^{2}r^{4-2D}}{1-24\alpha \Lambda }\right)}{ 3\left(D-3\right) \left( 2D-5\right) \left( D-1\right) \sqrt{r^{4D}\left(1-24\alpha \Lambda -6\alpha Q^{2}r^{4-2D}\right)}}\Bigg\}
\nonumber
\\
&+\textrm{constant}\Bigg)
\end{align}
where $Q$ and $\alpha$ are arbitrary constants, while $\Lambda$ is the cosmological constant. Since we are interested in asymptotically flat spacetime, we require $\Lambda =0$. This leads to much simpler form of the above metric elements,
\begin{align}
T(r)&=\frac{-1\pm \sqrt{1-6\alpha Q^{2}r^{4-2D}}}{6\alpha}
\end{align}
\begin{align}
g(r)&=\left(\frac{2}{3}\pm\frac{1}{3}\sqrt{1-6\alpha Q^{2}r^{4-2D}}\right) ^{-2}r^{3-D}
\Bigg( \frac{1}{\left( D-2\right) }
 \Bigg\{\frac{1}{54\alpha} \Big[-\frac{18\alpha Q^{2}r^{3-D}}{3-D}-\frac{r^{D-1}}{D-1}\Big]
\nonumber
\\
&\pm \frac{\sqrt{r^{4D}\left(1-6\alpha
Q^{2}r^{4-2D}\right) }}{54\alpha}\Big[ \frac{6\alpha
Q^{2}r^{3-3D}}{2D-5}+\frac{ r^{-1-D}}{D-1}\Big]
\nonumber
\\
&\pm \frac{\left(D-2\right)^{2}Q^{2}r^{3+D}\sqrt{1-6\alpha Q^{2}r^{4-2D}}\ _2F_1\left(
\frac{D-3}{2\left( D-2\right) },\frac{1}{2},\frac{3D-7}{2\left( D-2\right)};6\alpha Q^{2}r^{4-2D}\right)}{ 3\left(D-3\right) \left( 2D-5\right) \left(D-1\right) \sqrt{r^{4D}\left(
1-6\alpha Q^{2}r^{4-2D}\right)}}\Bigg\}
\nonumber\\
&+\textrm{constants}\Bigg) 
\end{align}
\begin{align}
f(r)&=\frac{1}{r^{D-3}}\Bigg(\frac{1}{\left( D-2\right)}\Bigg\{ \frac{1}{54\alpha} 
\Big[-\frac{18\alpha Q^{2}r^{3-D}}{3-D}-\frac{r^{D-1}}{D-1}\Big]
\nonumber
\\
&\pm \frac{\sqrt{r^{4D}\left(1-6\alpha Q^{2}r^{4-2D}\right) }}{54\alpha}\Big[\frac{6\alpha
Q^{2}r^{3-3D}}{2D-5}+\frac{r^{-1-D}}{D-1}\Big]
\nonumber
\\
&\pm \frac{\left(D-2\right) ^{2}Q^{2}r^{3+D}\sqrt{1-6\alpha Q^{2}r^{4-2D}}\ _2F_1\left(
\frac{D-3}{2\left( D-2\right) },\frac{1}{2},\frac{3D-7}{2\left( D-2\right)};
6\alpha Q^{2}r^{4-2D}\right)}{ 3\left(D-3\right) \left( 2D-5\right) \left( D-1\right) \sqrt{r^{4D}\left(1-6\alpha Q^{2}r^{4-2D}\right)}}\Bigg\}
\nonumber
\\
&+\textrm{constant}\Bigg)
\end{align}
Further, taking the limit $\alpha \rightarrow 0$, with the upper sign in the torsional part, we obtain, $T(r)=-\tilde{q}/r^{4}$. Here $\tilde{q}=Q^{2}/2$ and is the torsion parameter present in the model. Under the same limit, the metric elements reduce to those given in \ref{Eq_Torso_Rev}.

\bibliography{Gravity_1_full,Gravity_2_partial,Brane,My_References}

\providecommand{\href}[2]{#2}\begingroup\raggedright\begin{thebibliography}{10}

\bibitem{Schneider1992}
P.~{Schneider}, J.~{Ehlers}, and E.~E. {Falco},
  \href{http://dx.doi.org/10.1007/978-3-662-03758-4}{{\em {Gravitational
  Lenses}}}.
\newblock Springer-Verlag, 1992.

\bibitem{Liebes:1964zz}
S.~Liebes, ``{Gravitational Lenses},''
\href{http://dx.doi.org/10.1103/PhysRev.133.B835}{{\em Phys. Rev.} {\bfseries
  133} (1964) B835--B844}.

\bibitem{Refsdal:1993kf}
S.~Refsdal and J.~Surdej, ``{Gravitational lenses},''
\href{http://dx.doi.org/10.1088/0034-4885/57/2/001}{{\em Rept. Prog. Phys.}
  {\bfseries 57} (1994) 117--186}.

\bibitem{Darwin:1959}
C.~Darwin, ``{The gravity field of a particle},''
  \href{http://dx.doi.org/10.1098/rspa.1959.0015}{{\em Proc. Roy. Soc. Lond. A}
  {\bfseries 249} (1959) 180}.

\bibitem{Ohanian:1987}
H.~Ohanian, ``{The black hole as a gravitational ``lens"},''
  \href{http://dx.doi.org/10.1119/1.15126}{{\em Am. J. Phys.} {\bfseries 55}
  (1987) 428}.

\bibitem{Luminet:1979}
J.-P. Luminet, ``{Image of a spherical black hole with thin accretion disk},''
  \href{http://dx.doi.org/1979A%26A....75..228L}{{\em Astron. Astroph.}
  {\bfseries 75} (May, 1979) 228--235}.

\bibitem{Virbhadra:1999nm}
K.~S. Virbhadra and G.~F.~R. Ellis, ``{Schwarzschild black hole lensing},''
  \href{http://dx.doi.org/10.1103/PhysRevD.62.084003}{{\em Phys. Rev.}
  {\bfseries D62} (2000) 084003},
\href{http://arxiv.org/abs/astro-ph/9904193}{{\ttfamily arXiv:astro-ph/9904193
  [astro-ph]}}.

\bibitem{Frittelli:1999yf}
S.~Frittelli, T.~P. Kling, and E.~T. Newman, ``{Space-time perspective of
  Schwarzschild lensing},''
  \href{http://dx.doi.org/10.1103/PhysRevD.61.064021}{{\em Phys. Rev.}
  {\bfseries D61} (2000) 064021},
\href{http://arxiv.org/abs/gr-qc/0001037}{{\ttfamily arXiv:gr-qc/0001037
  [gr-qc]}}.

\bibitem{Bozza:2001xd}
V.~Bozza, S.~Capozziello, G.~Iovane, and G.~Scarpetta, ``{Strong field limit of
  black hole gravitational lensing},''
  \href{http://dx.doi.org/10.1023/A:1012292927358}{{\em Gen. Rel. Grav.}
  {\bfseries 33} (2001) 1535--1548},
\href{http://arxiv.org/abs/gr-qc/0102068}{{\ttfamily arXiv:gr-qc/0102068
  [gr-qc]}}.

\bibitem{Bozza:2002zj}
V.~Bozza, ``{Gravitational lensing in the strong field limit},''
  \href{http://dx.doi.org/10.1103/PhysRevD.66.103001}{{\em Phys. Rev.}
  {\bfseries D66} (2002) 103001},
\href{http://arxiv.org/abs/gr-qc/0208075}{{\ttfamily arXiv:gr-qc/0208075
  [gr-qc]}}.

\bibitem{Bozza:2007gt}
V.~Bozza and G.~Scarpetta, ``{Strong deflection limit of black hole
  gravitational lensing with arbitrary source distances},''
  \href{http://dx.doi.org/10.1103/PhysRevD.76.083008}{{\em Phys. Rev.}
  {\bfseries D76} (2007) 083008},
\href{http://arxiv.org/abs/0705.0246}{{\ttfamily arXiv:0705.0246 [gr-qc]}}.

\bibitem{Eiroa:2002mk}
E.~F. Eiroa, G.~E. Romero, and D.~F. Torres, ``{Reissner-Nordstrom black hole
  lensing},'' \href{http://dx.doi.org/10.1103/PhysRevD.66.024010}{{\em Phys.
  Rev.} {\bfseries D66} (2002) 024010},
\href{http://arxiv.org/abs/gr-qc/0203049}{{\ttfamily arXiv:gr-qc/0203049
  [gr-qc]}}.

\bibitem{Bozza:2006nm}
V.~Bozza, F.~De~Luca, and G.~Scarpetta, ``{Kerr black hole lensing for generic
  observers in the strong deflection limit},''
  \href{http://dx.doi.org/10.1103/PhysRevD.74.063001}{{\em Phys. Rev.}
  {\bfseries D74} (2006) 063001},
\href{http://arxiv.org/abs/gr-qc/0604093}{{\ttfamily arXiv:gr-qc/0604093
  [gr-qc]}}.

\bibitem{Kraniotis:2014paa}
G.~V. Kraniotis, ``{Gravitational lensing and frame dragging of light in the
  Kerr-Newman and the Kerr-Newman-(anti) de Sitter black hole spacetimes},''
  \href{http://dx.doi.org/10.1007/s10714-014-1818-8}{{\em Gen. Rel. Grav.}
  {\bfseries 46} no.~11, (2014) 1818},
\href{http://arxiv.org/abs/1401.7118}{{\ttfamily arXiv:1401.7118 [gr-qc]}}.

\bibitem{Eiroa:2005ag}
E.~F. Eiroa, ``{Gravitational lensing by Einstein-Born-Infeld black holes},''
  \href{http://dx.doi.org/10.1103/PhysRevD.73.043002}{{\em Phys. Rev.}
  {\bfseries D73} (2006) 043002},
\href{http://arxiv.org/abs/gr-qc/0511065}{{\ttfamily arXiv:gr-qc/0511065
  [gr-qc]}}.

\bibitem{Chen:2009eu}
S.-b. Chen and J.-l. Jing, ``{Strong field gravitational lensing in the
  deformed Horava-Lifshitz black hole},''
  \href{http://dx.doi.org/10.1103/PhysRevD.80.024036}{{\em Phys. Rev.}
  {\bfseries D80} (2009) 024036},
\href{http://arxiv.org/abs/0905.2055}{{\ttfamily arXiv:0905.2055 [gr-qc]}}.

\bibitem{Whisker:2004gq}
R.~Whisker, ``{Strong gravitational lensing by braneworld black holes},''
  \href{http://dx.doi.org/10.1103/PhysRevD.71.064004}{{\em Phys. Rev.}
  {\bfseries D71} (2005) 064004},
\href{http://arxiv.org/abs/astro-ph/0411786}{{\ttfamily arXiv:astro-ph/0411786
  [astro-ph]}}.

\bibitem{Bhadra:2003zs}
A.~Bhadra, ``{Gravitational lensing by a charged black hole of string
  theory},'' \href{http://dx.doi.org/10.1103/PhysRevD.67.103009}{{\em Phys.
  Rev.} {\bfseries D67} (2003) 103009},
\href{http://arxiv.org/abs/gr-qc/0306016}{{\ttfamily arXiv:gr-qc/0306016
  [gr-qc]}}.

\bibitem{Ghosh:2010uw}
T.~Ghosh and S.~Sengupta, ``{Strong gravitational lensing across dilaton
  anti-de Sitter black hole},''
  \href{http://dx.doi.org/10.1103/PhysRevD.81.044013}{{\em Phys. Rev.}
  {\bfseries D81} (2010) 044013},
\href{http://arxiv.org/abs/1001.5129}{{\ttfamily arXiv:1001.5129 [gr-qc]}}.

\bibitem{Mukherjee:2006ru}
N.~Mukherjee and A.~S. Majumdar, ``{Particle motion and gravitational lensing
  in the metric of a dilaton black hole in a de Sitter universe},''
  \href{http://dx.doi.org/10.1007/s10714-007-0407-5}{{\em Gen. Rel. Grav.}
  {\bfseries 39} (2007) 583--600},
\href{http://arxiv.org/abs/astro-ph/0605224}{{\ttfamily arXiv:astro-ph/0605224
  [astro-ph]}}.

\bibitem{Eiroa:2013nra}
E.~F. Eiroa and C.~M. Sendra, ``{Regular phantom black hole gravitational
  lensing},'' \href{http://dx.doi.org/10.1103/PhysRevD.88.103007}{{\em Phys.
  Rev.} {\bfseries D88} no.~10, (2013) 103007},
\href{http://arxiv.org/abs/1308.5959}{{\ttfamily arXiv:1308.5959 [gr-qc]}}.

\bibitem{Gyulchev:2012ty}
G.~N. Gyulchev and I.~Z. Stefanov, ``{Gravitational Lensing by Phantom Black
  holes},'' \href{http://dx.doi.org/10.1103/PhysRevD.87.063005}{{\em Phys.
  Rev.} {\bfseries D87} no.~6, (2013) 063005},
\href{http://arxiv.org/abs/1211.3458}{{\ttfamily arXiv:1211.3458 [gr-qc]}}.

\bibitem{Zhao:2016kft}
S.-S. Zhao and Y.~Xie, ``{Strong field gravitational lensing by a charged
  Galileon black hole},''
  \href{http://dx.doi.org/10.1088/1475-7516/2016/07/007}{{\em JCAP} {\bfseries
  1607} no.~07, (2016) 007},
\href{http://arxiv.org/abs/1603.00637}{{\ttfamily arXiv:1603.00637 [gr-qc]}}.

\bibitem{Nandi:2006ds}
K.~K. Nandi, Y.-Z. Zhang, and A.~V. Zakharov, ``{Gravitational lensing by
  wormholes},'' \href{http://dx.doi.org/10.1103/PhysRevD.74.024020}{{\em Phys.
  Rev.} {\bfseries D74} (2006) 024020},
\href{http://arxiv.org/abs/gr-qc/0602062}{{\ttfamily arXiv:gr-qc/0602062
  [gr-qc]}}.

\bibitem{Wei:2015qca}
S.~W. Wei, Y.~X. Liu, and C.~E. Fu, ``{Null Geodesics and Gravitational Lensing
  in a Nonsingular Spacetime},''
  \href{http://dx.doi.org/10.1155/2015/454217}{{\em Adv. High Energy Phys.}
  {\bfseries 2015} (2015) 454217},
\href{http://arxiv.org/abs/1510.02560}{{\ttfamily arXiv:1510.02560 [gr-qc]}}.

\bibitem{Sahu:2015dea}
S.~Sahu, K.~Lochan, and D.~Narasimha, ``{Gravitational lensing by self-dual
  black holes in loop quantum gravity},''
  \href{http://dx.doi.org/10.1103/PhysRevD.91.063001}{{\em Phys. Rev.}
  {\bfseries D91} (2015) 063001},
\href{http://arxiv.org/abs/1502.05619}{{\ttfamily arXiv:1502.05619 [gr-qc]}}.

\bibitem{Virbhadra:2002ju}
K.~S. Virbhadra and G.~F.~R. Ellis, ``{Gravitational lensing by naked
  singularities},''
\href{http://dx.doi.org/10.1103/PhysRevD.65.103004}{{\em Phys. Rev.} {\bfseries
  D65} (2002) 103004}.

\bibitem{Sahu:2012er}
S.~Sahu, M.~Patil, D.~Narasimha, and P.~S. Joshi, ``{Can strong gravitational
  lensing distinguish naked singularities from black holes?},''
  \href{http://dx.doi.org/10.1103/PhysRevD.86.063010}{{\em Phys. Rev.}
  {\bfseries D86} (2012) 063010},
\href{http://arxiv.org/abs/1206.3077}{{\ttfamily arXiv:1206.3077 [gr-qc]}}.

\bibitem{Amore:2006mc}
P.~Amore, M.~Cervantes, A.~De~Pace, and F.~M. Fernandez, ``{Gravitational
  lensing from compact bodies: Analytical results for strong and weak
  deflection limits},''
  \href{http://dx.doi.org/10.1103/PhysRevD.75.083005}{{\em Phys. Rev.}
  {\bfseries D75} (2007) 083005},
\href{http://arxiv.org/abs/gr-qc/0610153}{{\ttfamily arXiv:gr-qc/0610153
  [gr-qc]}}.

\bibitem{Bozza:2009yw}
V.~Bozza, ``{Gravitational Lensing by Black Holes},''
  \href{http://dx.doi.org/10.1007/s10714-010-0988-2}{{\em Gen. Rel. Grav.}
  {\bfseries 42} (2010) 2269--2300},
\href{http://arxiv.org/abs/0911.2187}{{\ttfamily arXiv:0911.2187 [gr-qc]}}.

\bibitem{Raffaelli:2014ola}
B.~Raffaelli, ``{Strong gravitational lensing and black hole quasinormal modes:
  Towards a semiclassical unified description},''
  \href{http://dx.doi.org/10.1007/s10714-016-2016-7}{{\em Gen. Rel. Grav.}
  {\bfseries 48} no.~2, (2016) 16},
\href{http://arxiv.org/abs/1412.7333}{{\ttfamily arXiv:1412.7333 [gr-qc]}}.

\bibitem{Loeb:2013lfa}
{\bfseries Perimeter Institute for Theoretical Physics} Collaboration, A.~E.
  Broderick, T.~Johannsen, A.~Loeb, and D.~Psaltis, ``{Testing the No-Hair
  Theorem with Event Horizon Telescope Observations of Sagittarius A*},''
  \href{http://dx.doi.org/10.1088/0004-637X/784/1/7}{{\em Astrophys. J.}
  {\bfseries 784} (2014) 7},
\href{http://arxiv.org/abs/1311.5564}{{\ttfamily arXiv:1311.5564
  [astro-ph.HE]}}.

\bibitem{Bozza:2004kq}
V.~Bozza and L.~Mancini, ``{Gravitational lensing by black holes: A
  Comprehensive treatment and the case of the star S2},''
  \href{http://dx.doi.org/10.1086/422309}{{\em Astrophys. J.} {\bfseries 611}
  (2004) 1045--1053},
\href{http://arxiv.org/abs/astro-ph/0404526}{{\ttfamily arXiv:astro-ph/0404526
  [astro-ph]}}.

\bibitem{Abdujabbarov:2015xqa}
A.~A. Abdujabbarov, L.~Rezzolla, and B.~J. Ahmedov, ``{A coordinate-independent
  characterization of a black hole shadow},''
  \href{http://dx.doi.org/10.1093/mnras/stv2079}{{\em Mon. Not. Roy. Astron.
  Soc.} {\bfseries 454} no.~3, (2015) 2423--2435},
\href{http://arxiv.org/abs/1503.09054}{{\ttfamily arXiv:1503.09054 [gr-qc]}}.

\bibitem{Falcke:2013ola}
H.~Falcke and S.~B. Markoff, ``{Toward the event horizon—the supermassive
  black hole in the Galactic Center},''
  \href{http://dx.doi.org/10.1088/0264-9381/30/24/244003}{{\em Class. Quant.
  Grav.} {\bfseries 30} (2013) 244003},
\href{http://arxiv.org/abs/1311.1841}{{\ttfamily arXiv:1311.1841
  [astro-ph.HE]}}.

\bibitem{Bozza:2012by}
V.~Bozza and L.~Mancini, ``{Observing gravitational lensing effects by Sgr A*
  with GRAVITY},'' \href{http://dx.doi.org/10.1088/0004-637X/753/1/56}{{\em
  Astrophys. J.} {\bfseries 753} (2012) 56},
\href{http://arxiv.org/abs/1204.2103}{{\ttfamily arXiv:1204.2103
  [astro-ph.GA]}}.

\bibitem{BinNun:2010ty}
A.~Y. Bin-Nun, ``{Strong Gravitational Lensing by Sgr A*},''
  \href{http://dx.doi.org/10.1088/0264-9381/28/11/114003}{{\em Class. Quant.
  Grav.} {\bfseries 28} (2011) 114003},
\href{http://arxiv.org/abs/1011.5848}{{\ttfamily arXiv:1011.5848 [gr-qc]}}.

\bibitem{Bozza:2015wbw}
V.~Bozza and C.~Melchiorre, ``{Caustics of $1/r^n$ binary gravitational lenses:
  from galactic haloes to exotic matter},''
  \href{http://dx.doi.org/10.1088/1475-7516/2016/03/040}{{\em JCAP} {\bfseries
  1603} no.~03, (2016) 040},
\href{http://arxiv.org/abs/1511.07991}{{\ttfamily arXiv:1511.07991 [gr-qc]}}.

\bibitem{Bozza:2007gm}
V.~Bozza, S.~C. Novati, and L.~Mancini, ``{Gravitational lensing by the
  supermassive black hole in the center of M31},''
  \href{http://dx.doi.org/10.1393/ncb/i2007-10386-6}{{\em Astrophys. J.}
  {\bfseries 675} (2008) 340}, \href{http://arxiv.org/abs/0711.0750}{{\ttfamily
  arXiv:0711.0750 [astro-ph]}}.
[Nuovo Cim.B122,579(2007)].

\bibitem{Bozza:2014ywa}
V.~Bozza, ``{Gravitational lensing by black holes: The case of Sgr A$^*$},''
\href{http://dx.doi.org/10.1063/1.4861946}{{\em AIP Conf. Proc.} {\bfseries
  1577} (2014) 89--93}.

\bibitem{Bozza:2015haa}
V.~Bozza and A.~Postiglione, ``{Alternatives to Schwarzschild in the weak field
  limit of General Relativity},''
  \href{http://dx.doi.org/10.1088/1475-7516/2015/06/036}{{\em JCAP} {\bfseries
  1506} no.~06, (2015) 036},
\href{http://arxiv.org/abs/1502.05178}{{\ttfamily arXiv:1502.05178 [gr-qc]}}.

\bibitem{Aldi:2016ntn}
G.~F. Aldi and V.~Bozza, ``{Relativistic iron lines in accretion disks: the
  contribution of higher order images in the strong deflection limit},''
\href{http://arxiv.org/abs/1607.05365}{{\ttfamily arXiv:1607.05365
  [astro-ph.HE]}}.

\bibitem{Green:1987mn}
M.~B. Green, J.~H. Schwarz, and E.~Witten, {\em {Superstring Theory. Vol. 2:
  Loop Amplitudes, Anomalies and Phenomenology}}.
\newblock 1988.
\newblock
\url{http://www.cambridge.org/us/academic/subjects/physics/theoretical-physics-and-mathematical-physics/superstring-theory-volume-2}.
\newblock

\bibitem{Chakraborty:2014xla}
S.~Chakraborty and S.~SenGupta, ``{Spherically symmetric brane spacetime with
  bulk $f(\mathcal {R})$ gravity},''
  \href{http://dx.doi.org/10.1140/epjc/s10052-014-3234-3}{{\em Eur.Phys.J.}
  {\bfseries C75} no.~1, (2015) 11},
\href{http://arxiv.org/abs/1409.4115}{{\ttfamily arXiv:1409.4115 [gr-qc]}}.

\bibitem{Chakraborty:2015bja}
S.~Chakraborty and S.~SenGupta, ``{Effective gravitational field equations on
  $m$-brane embedded in n-dimensional bulk of Einstein and $f(\mathcal {R})$
  gravity},'' \href{http://dx.doi.org/10.1140/epjc/s10052-015-3768-z}{{\em Eur.
  Phys. J.} {\bfseries C75} no.~11, (2015) 538},
\href{http://arxiv.org/abs/1504.07519}{{\ttfamily arXiv:1504.07519 [gr-qc]}}.

\bibitem{Chakraborty:2015taq}
S.~Chakraborty and S.~SenGupta, ``{Spherically symmetric brane in a bulk of
  f(R) and Gauss-Bonnet Gravity},''
  \href{http://dx.doi.org/10.1088/0264-9381/33/22/225001}{{\em Class. Quant.
  Grav.} {\bfseries 33} no.~22, (2016) 225001},
\href{http://arxiv.org/abs/1510.01953}{{\ttfamily arXiv:1510.01953 [gr-qc]}}.

\bibitem{Chakraborty:2016ydo}
S.~Chakraborty and S.~SenGupta, ``{Solving higher curvature gravity
  theories},'' \href{http://dx.doi.org/10.1140/epjc/s10052-016-4394-0}{{\em
  Eur. Phys. J.} {\bfseries C76} no.~10, (2016) 552},
\href{http://arxiv.org/abs/1604.05301}{{\ttfamily arXiv:1604.05301 [gr-qc]}}.

\bibitem{gravitation}
T.Padmanabhan, {\em {Gravitation: Foundations and Frontiers}}.
\newblock Cambridge University Press, Cambridge, UK, 2010.

\bibitem{Howe:1996kj}
P.~S. Howe and G.~Papadopoulos, ``{Twistor spaces for HKT manifolds},''
  \href{http://dx.doi.org/10.1016/0370-2693(96)00393-0}{{\em Phys. Lett.}
  {\bfseries B379} (1996) 80--86},
\href{http://arxiv.org/abs/hep-th/9602108}{{\ttfamily arXiv:hep-th/9602108
  [hep-th]}}.

\bibitem{Kar:2001eb}
S.~Kar, P.~Majumdar, S.~SenGupta, and S.~Sur, ``{Cosmic optical activity from
  an inhomogeneous Kalb-Ramond field},''
  \href{http://dx.doi.org/10.1088/0264-9381/19/4/304}{{\em Class. Quant. Grav.}
  {\bfseries 19} (2002) 677--688},
\href{http://arxiv.org/abs/hep-th/0109135}{{\ttfamily arXiv:hep-th/0109135
  [hep-th]}}.

\bibitem{Letelier:1995ze}
P.~S. Letelier, ``{Spinning strings as torsion line space-time defects},''
\href{http://dx.doi.org/10.1088/0264-9381/12/2/016}{{\em Class. Quant. Grav.}
  {\bfseries 12} (1995) 471--478}.

\bibitem{Ellis:2013gca}
J.~Ellis, N.~E. Mavromatos, and S.~Sarkar, ``{Environmental CPT Violation in an
  Expanding Universe in String Theory},''
  \href{http://dx.doi.org/10.1016/j.physletb.2013.07.016}{{\em Phys. Lett.}
  {\bfseries B725} (2013) 407--411},
\href{http://arxiv.org/abs/1304.5433}{{\ttfamily arXiv:1304.5433 [gr-qc]}}.

\bibitem{Maity:2004he}
D.~Maity, P.~Majumdar, and S.~SenGupta, ``{Parity violating Kalb-Ramond-Maxwell
  interactions and CMB anisotropy in a brane world},''
  \href{http://dx.doi.org/10.1088/1475-7516/2004/06/005}{{\em JCAP} {\bfseries
  0406} (2004) 005},
\href{http://arxiv.org/abs/hep-th/0401218}{{\ttfamily arXiv:hep-th/0401218
  [hep-th]}}.

\bibitem{Majumdar:1999jd}
P.~Majumdar and S.~SenGupta, ``{Parity violating gravitational coupling of
  electromagnetic fields},''
  \href{http://dx.doi.org/10.1088/0264-9381/16/12/102}{{\em Class. Quant.
  Grav.} {\bfseries 16} (1999) L89--L94},
\href{http://arxiv.org/abs/gr-qc/9906027}{{\ttfamily arXiv:gr-qc/9906027
  [gr-qc]}}.

\bibitem{Myrzakulov:2010vz}
R.~Myrzakulov, ``{Accelerating universe from F(T) gravity},''
  \href{http://dx.doi.org/10.1140/epjc/s10052-011-1752-9}{{\em Eur. Phys. J.}
  {\bfseries C71} (2011) 1752},
\href{http://arxiv.org/abs/1006.1120}{{\ttfamily arXiv:1006.1120 [gr-qc]}}.

\bibitem{Chen:2010va}
S.-H. Chen, J.~B. Dent, S.~Dutta, and E.~N. Saridakis, ``{Cosmological
  perturbations in f(T) gravity},''
  \href{http://dx.doi.org/10.1103/PhysRevD.83.023508}{{\em Phys. Rev.}
  {\bfseries D83} (2011) 023508},
\href{http://arxiv.org/abs/1008.1250}{{\ttfamily arXiv:1008.1250
  [astro-ph.CO]}}.

\bibitem{Dent:2011zz}
J.~B. Dent, S.~Dutta, and E.~N. Saridakis, ``{f(T) gravity mimicking dynamical
  dark energy. Background and perturbation analysis},''
  \href{http://dx.doi.org/10.1088/1475-7516/2011/01/009}{{\em JCAP} {\bfseries
  1101} (2011) 009},
\href{http://arxiv.org/abs/1010.2215}{{\ttfamily arXiv:1010.2215
  [astro-ph.CO]}}.

\bibitem{Bamba:2010wb}
K.~Bamba, C.-Q. Geng, C.-C. Lee, and L.-W. Luo, ``{Equation of state for dark
  energy in $f(T)$ gravity},''
  \href{http://dx.doi.org/10.1088/1475-7516/2011/01/021}{{\em JCAP} {\bfseries
  1101} (2011) 021},
\href{http://arxiv.org/abs/1011.0508}{{\ttfamily arXiv:1011.0508
  [astro-ph.CO]}}.

\bibitem{Zhang:2011qp}
Y.~Zhang, H.~Li, Y.~Gong, and Z.-H. Zhu, ``{Notes on $f(T)$ Theories},''
  \href{http://dx.doi.org/10.1088/1475-7516/2011/07/015}{{\em JCAP} {\bfseries
  1107} (2011) 015},
\href{http://arxiv.org/abs/1103.0719}{{\ttfamily arXiv:1103.0719
  [astro-ph.CO]}}.

\bibitem{Chakraborty:2012kj}
S.~Chakraborty, ``{An alternative f(R, T) gravity theory and the dark energy
  problem},'' \href{http://dx.doi.org/10.1007/s10714-013-1577-y}{{\em Gen. Rel.
  Grav.} {\bfseries 45} (2013) 2039--2052},
\href{http://arxiv.org/abs/1212.3050}{{\ttfamily arXiv:1212.3050
  [physics.gen-ph]}}.

\bibitem{Yang:2010ji}
R.-J. Yang, ``{Conformal transformation in $f(T)$ theories},''
  \href{http://dx.doi.org/10.1209/0295-5075/93/60001}{{\em Europhys. Lett.}
  {\bfseries 93} (2011) 60001},
\href{http://arxiv.org/abs/1010.1376}{{\ttfamily arXiv:1010.1376 [gr-qc]}}.

\bibitem{Bhattacharya:2016naa}
S.~Bhattacharya and S.~Chakraborty, ``{Solar system constraints on some
  Horndeski gravity theories},''
\href{http://arxiv.org/abs/1607.03693}{{\ttfamily arXiv:1607.03693 [gr-qc]}}.

\bibitem{Chakraborty:2011uj}
S.~Chakraborty and S.~Chakraborty, ``{Trajectory around a spherically symmetric
  non-rotating black hole},'' \href{http://dx.doi.org/10.1139/p11-032}{{\em
  Can.J.Phys.} {\bfseries 89} (2011) 689--695},
\href{http://arxiv.org/abs/1109.0676}{{\ttfamily arXiv:1109.0676 [gr-qc]}}.

\bibitem{Chakraborty:2014eha}
S.~Chakraborty, ``{Equilibrium configuration of perfect fluid orbiting around
  black holes in some classes of alternative gravity theories},''
  \href{http://dx.doi.org/10.1088/0264-9381/32/7/075007}{{\em
  Class.Quant.Grav.} {\bfseries 32} no.~7, (2015) 075007},
\href{http://arxiv.org/abs/1406.0417}{{\ttfamily arXiv:1406.0417 [gr-qc]}}.

\bibitem{Chakraborty:2015vla}
S.~Chakraborty, ``{Aspects of Neutrino Oscillation in Alternative Gravity
  Theories},'' \href{http://dx.doi.org/10.1088/1475-7516/2015/10/019}{{\em
  JCAP} {\bfseries 1510} no.~10, (2015) 019},
\href{http://arxiv.org/abs/1506.02647}{{\ttfamily arXiv:1506.02647 [gr-qc]}}.

\bibitem{Bozza:2008ev}
V.~Bozza, ``{A Comparison of approximate gravitational lens equations and a
  proposal for an improved new one},''
  \href{http://dx.doi.org/10.1103/PhysRevD.78.103005}{{\em Phys. Rev.}
  {\bfseries D78} (2008) 103005},
\href{http://arxiv.org/abs/0807.3872}{{\ttfamily arXiv:0807.3872 [gr-qc]}}.

\bibitem{Kar:2002xa}
S.~Kar, S.~SenGupta, and S.~Sur, ``{Static spherisymmetric solutions,
  gravitational lensing and perihelion precession in Einstein-Kalb-Ramond
  theory},'' \href{http://dx.doi.org/10.1103/PhysRevD.67.044005}{{\em Phys.
  Rev.} {\bfseries D67} (2003) 044005},
\href{http://arxiv.org/abs/hep-th/0210176}{{\ttfamily arXiv:hep-th/0210176
  [hep-th]}}.

\bibitem{SenGupta:2001cs}
S.~SenGupta and S.~Sur, ``{Spherically symmetric solutions of gravitational
  field equations in Kalb-Ramond background},''
  \href{http://dx.doi.org/10.1016/S0370-2693(01)01238-2}{{\em Phys. Lett.}
  {\bfseries B521} (2001) 350--356},
\href{http://arxiv.org/abs/gr-qc/0102095}{{\ttfamily arXiv:gr-qc/0102095
  [gr-qc]}}.

\bibitem{Chakraborty:2014fva}
S.~Chakraborty and S.~SenGupta, ``{Solutions on a brane in a bulk spacetime
  with Kalb-Ramond field},''
  \href{http://dx.doi.org/10.1016/j.aop.2016.01.023}{{\em Annals Phys.}
  {\bfseries 367} (2016) 258--279},
\href{http://arxiv.org/abs/1412.7783}{{\ttfamily arXiv:1412.7783 [gr-qc]}}.

\bibitem{Dadhich:2000am}
N.~Dadhich, R.~Maartens, P.~Papadopoulos, and V.~Rezania, ``{Black holes on the
  brane},'' \href{http://dx.doi.org/10.1016/S0370-2693(00)00798-X}{{\em
  Phys.Lett.} {\bfseries B487} (2000) 1--6},
\href{http://arxiv.org/abs/hep-th/0003061}{{\ttfamily arXiv:hep-th/0003061
  [hep-th]}}.

\bibitem{Ferraro:2006jd}
R.~Ferraro and F.~Fiorini, ``{Modified teleparallel gravity: Inflation without
  inflaton},'' \href{http://dx.doi.org/10.1103/PhysRevD.75.084031}{{\em Phys.
  Rev.} {\bfseries D75} (2007) 084031},
\href{http://arxiv.org/abs/gr-qc/0610067}{{\ttfamily arXiv:gr-qc/0610067
  [gr-qc]}}.

\bibitem{Ferraro:2008ey}
R.~Ferraro and F.~Fiorini, ``{On Born-Infeld Gravity in Weitzenbock
  spacetime},'' \href{http://dx.doi.org/10.1103/PhysRevD.78.124019}{{\em Phys.
  Rev.} {\bfseries D78} (2008) 124019},
\href{http://arxiv.org/abs/0812.1981}{{\ttfamily arXiv:0812.1981 [gr-qc]}}.

\bibitem{Daouda:2011rt}
M.~Hamani~Daouda, M.~E. Rodrigues, and M.~J.~S. Houndjo, ``{Static Anisotropic
  Solutions in f(T) Theory},''
  \href{http://dx.doi.org/10.1140/epjc/s10052-012-1890-8}{{\em Eur. Phys. J.}
  {\bfseries C72} (2012) 1890},
\href{http://arxiv.org/abs/1109.0528}{{\ttfamily arXiv:1109.0528
  [physics.gen-ph]}}.

\bibitem{Capozziello:2001mq}
S.~Capozziello, G.~Lambiase, and C.~Stornaiolo, ``{Geometric classification of
  the torsion tensor in space-time},''
  \href{http://dx.doi.org/10.1002/1521-3889(200108)10:8<713::AID-ANDP713>3.0.CO,
  10.1002/1521-3889(200108)10:8<713::AID-ANDP713>3.0.CO;2-2}{{\em Annalen
  Phys.} {\bfseries 10} (2001) 713--727},
\href{http://arxiv.org/abs/gr-qc/0101038}{{\ttfamily arXiv:gr-qc/0101038
  [gr-qc]}}.

\bibitem{Capozziello:2007ec}
S.~Capozziello and M.~Francaviglia, ``{Extended Theories of Gravity and their
  Cosmological and Astrophysical Applications},''
  \href{http://dx.doi.org/10.1007/s10714-007-0551-y}{{\em Gen. Rel. Grav.}
  {\bfseries 40} (2008) 357--420},
\href{http://arxiv.org/abs/0706.1146}{{\ttfamily arXiv:0706.1146 [astro-ph]}}.

\bibitem{Myrzakulov:2012qp}
R.~Myrzakulov, ``{FRW Cosmology in F(R,T) gravity},''
  \href{http://dx.doi.org/10.1140/epjc/s10052-012-2203-y}{{\em Eur. Phys. J.}
  {\bfseries C72} (2012) 2203},
\href{http://arxiv.org/abs/1207.1039}{{\ttfamily arXiv:1207.1039 [gr-qc]}}.

\bibitem{Wanas:2012pu}
M.~I. Wanas and H.~A. Hassan, ``{Torsion and Problems of Standard Cosmology},''
\href{http://arxiv.org/abs/1209.6218}{{\ttfamily arXiv:1209.6218 [gr-qc]}}.

\bibitem{Capozziello:2010zz}
V.~Faraoni and S.~Capozziello,
  \href{http://dx.doi.org/10.1007/978-94-007-0165-6}{{\em {Beyond Einstein
  Gravity}}}, vol.~170. 170 of {\em Fundamental Theories of Physics}.
\newblock Springer, Dordrecht, 2011.
\newblock
\url{http://www.springerlink.com/content/hl1805/#section=801705&page=1}.
\newblock

\bibitem{Ferraro:2011us}
R.~Ferraro and F.~Fiorini, ``{Non trivial frames for f(T) theories of gravity
  and beyond},'' \href{http://dx.doi.org/10.1016/j.physletb.2011.06.049}{{\em
  Phys. Lett.} {\bfseries B702} (2011) 75--80},
\href{http://arxiv.org/abs/1103.0824}{{\ttfamily arXiv:1103.0824 [gr-qc]}}.

\bibitem{Miao:2011ki}
R.-X. Miao, M.~Li, and Y.-G. Miao, ``{Violation of the first law of black hole
  thermodynamics in $f(T)$ gravity},''
  \href{http://dx.doi.org/10.1088/1475-7516/2011/11/033}{{\em JCAP} {\bfseries
  1111} (2011) 033},
\href{http://arxiv.org/abs/1107.0515}{{\ttfamily arXiv:1107.0515 [hep-th]}}.

\bibitem{Lochan:2015bha}
K.~Lochan and S.~Chakraborty, ``{Discrete quantum spectrum of black holes},''
  \href{http://dx.doi.org/10.1016/j.physletb.2016.01.060}{{\em Phys. Lett.}
  {\bfseries B755} (2016) 37--42},
\href{http://arxiv.org/abs/1509.09010}{{\ttfamily arXiv:1509.09010 [gr-qc]}}.

\bibitem{Capozziello:2012zj}
S.~Capozziello, P.~A. Gonzalez, E.~N. Saridakis, and Y.~Vasquez, ``{Exact
  charged black-hole solutions in D-dimensional f(T) gravity: torsion vs
  curvature analysis},'' \href{http://dx.doi.org/10.1007/JHEP02(2013)039}{{\em
  JHEP} {\bfseries 02} (2013) 039},
\href{http://arxiv.org/abs/1210.1098}{{\ttfamily arXiv:1210.1098 [hep-th]}}.

\bibitem{Schodel:2002vg}
R.~Schodel {\em et~al.}, ``{A Star in a 15.2 year orbit around the supermassive
  black hole at the center of the Milky Way},''
  \href{http://dx.doi.org/10.1038/nature01121}{{\em Nature} (2002) },
  \href{http://arxiv.org/abs/astro-ph/0210426}{{\ttfamily
  arXiv:astro-ph/0210426 [astro-ph]}}.
[Nature419,694(2002)].

\bibitem{Schodel:2003gy}
R.~Schodel, T.~Ott, R.~Genzel, A.~Eckart, N.~Mouawad, and T.~Alexander,
  ``{Stellar dynamics in the central arcsecond of our Galaxy},''
  \href{http://dx.doi.org/10.1086/378122}{{\em Astrophys. J.} {\bfseries 596}
  (2003) 1015--1034},
\href{http://arxiv.org/abs/astro-ph/0306214}{{\ttfamily arXiv:astro-ph/0306214
  [astro-ph]}}.

\bibitem{Gillessen:2009ht}
S.~Gillessen, F.~Eisenhauer, T.~K. Fritz, H.~Bartko, K.~Dodds-Eden, O.~Pfuhl,
  T.~Ott, and R.~Genzel, ``{The orbit of the star S2 around SgrA* from VLT and
  Keck data},'' \href{http://dx.doi.org/10.1088/0004-637X/707/2/L114}{{\em
  Astrophys. J.} {\bfseries 707} (2009) L114--L117},
\href{http://arxiv.org/abs/0910.3069}{{\ttfamily arXiv:0910.3069
  [astro-ph.GA]}}.

\bibitem{Broderick:2009ph}
A.~E. Broderick, A.~Loeb, and R.~Narayan, ``{The Event Horizon of Sagittarius
  A*},'' \href{http://dx.doi.org/10.1088/0004-637X/701/2/1357}{{\em Astrophys.
  J.} {\bfseries 701} (2009) 1357--1366},
\href{http://arxiv.org/abs/0903.1105}{{\ttfamily arXiv:0903.1105
  [astro-ph.HE]}}.

\bibitem{Gillessen:2008qv}
S.~Gillessen, F.~Eisenhauer, S.~Trippe, T.~Alexander, R.~Genzel, F.~Martins,
  and T.~Ott, ``{Monitoring stellar orbits around the Massive Black Hole in the
  Galactic Center},''
  \href{http://dx.doi.org/10.1088/0004-637X/692/2/1075}{{\em Astrophys. J.}
  {\bfseries 692} (2009) 1075--1109},
\href{http://arxiv.org/abs/0810.4674}{{\ttfamily arXiv:0810.4674 [astro-ph]}}.

\bibitem{Ghez:2005dd}
A.~M. Ghez {\em et~al.}, ``{The first laser guide star adaptive optics
  observations of the galactic center: sgr a*'s infrared color and the extended
  red emission in its vicinity},'' \href{http://dx.doi.org/10.1086/497576}{{\em
  Astrophys. J.} {\bfseries 635} (2005) 1087--1094},
\href{http://arxiv.org/abs/astro-ph/0508664}{{\ttfamily arXiv:astro-ph/0508664
  [astro-ph]}}.

\bibitem{Weinberg:2004nj}
N.~N. Weinberg, M.~Milosavljevic, and A.~M. Ghez, ``{Stellar dynamics at the
  Galactic Center with a Thirty Meter Telescope},''
  \href{http://dx.doi.org/10.1086/428079}{{\em Astrophys. J.} {\bfseries 622}
  (2005) 878},
\href{http://arxiv.org/abs/astro-ph/0404407}{{\ttfamily arXiv:astro-ph/0404407
  [astro-ph]}}.

\bibitem{Bozza:2005sc}
V.~Bozza and L.~Mancini, ``{Gravitational lensing of stars in the central
  arcsecond of our Galaxy},'' \href{http://dx.doi.org/10.1086/430664}{{\em
  Astrophys. J.} {\bfseries 627} (2005) 790--802},
\href{http://arxiv.org/abs/astro-ph/0503664}{{\ttfamily arXiv:astro-ph/0503664
  [astro-ph]}}.

\end{thebibliography}\endgroup

\bibliographystyle{./utphys1}
\end{document}